\documentclass[hyper]{JHEP}
\pdfoutput=1
\usepackage{fixltx2e}
\usepackage{float}
\usepackage{graphicx,amssymb,bm,latexsym,amsmath}
\usepackage{epstopdf}
\usepackage{caption}

\title{Perturbations of spiky strings in flat spacetimes}
\author{Soumya Bhattacharya, Sayan Kar and  Kamal L. Panigrahi \\
Department of Physics and Centre for Theoretical Studies,\\Indian Institute of Technology Kharagpur,\\
Kharagpur-721 302, India\\
Email: \email{soumya557@cts.iitkgp.ernet.in, sayan,panigrahi@phy.iitkgp.ernet.in}}

\abstract{Perturbations of a class of semiclassical strings known
today as spiky strings, are studied using the well-known Jacobi
equations for small normal deformations of an embedded timelike
surface. It is shown that there exists {\em finite} normal
perturbations of the spiky string worldsheets embedded in a $2+1$
dimensional flat spacetime. Such perturbations lead to a rounding
off of the spikes, which, in a way, demonstrates the stable nature of
the unperturbed worldsheet. The same features appear for the dual
spiky string solution and in the spiky as well as their dual
solutions in $3+1$ dimensional flat spacetime. Our results are
based on exact solutions of the corresponding Jacobi equations
which we obtain and use while constructing the profiles of the
perturbed configurations.}


\begin{document}

\section{Introduction}
Rigidly rotating strings were first proposed by Burden and Tassie
in 1982 \cite{BurdenTassie1: 1982, BurdenTassie2: 1982} and
explored further in the next few years in a series of
papers \cite{BurdenTassie3: 1984, Burden: 1985}. Later, Embacher
\cite{Embacher: 1992}  obtained the complete class of cosmic string
solutions in flat spacetime that undergo rigid rotation about a
fixed axis. Subsequent work, by various authors, on such string configurations
have appeared in \cite{rigid:1,rigid:2,rigid:3}.
More recently, in \cite{Burden: 2008},
Burden has come up with additional comments on the work reported in \cite{ogawa:2008}.

\noindent In the context of  AdS/CFT correspondence
\cite{Maldacena:1997re}, the spiky string solutions have
re-appeared, through the seminal work by Kruczenski
\cite{Kruczenski:2004wg}, as a potential gravity dual to higher
twist operators in string theory. The focus, in string theory,
however, has been on the solutions which are closed strings with
spikes (in cosmic string nomenclature these are known as cusps).
The `semiclassicality' of these solutions is  argued from the
evaluation of the energies and angular momenta of such spiky
string configurations \cite{Gubser:2002tv}.  From the
AdS$_5$/CFT$_4$ perspective, as mentioned above, the spiky strings
are special because they correspond to certain higher twist
operators in ${\cal N}$ =4 Supersymmetric Yang Mills (SYM) theory.
Further, these spiky strings have been found to be a special class
of the general rigidly rotating strings. In proving AdS/CFT
duality, the appearance of so-called ${\it integrability}$ 
at the level of the classical string worldsheet
\cite{Mandal:2002fs}, \cite{Bena:2003wd} as well as in N=4 gauge
theory \cite{Minahan:2002ve} has played a key role. The
semiclassical string states in the gravity side have been used to
look for suitable gauge theory operators on the boundary, in
establishing the duality. In this connection, Hoffman and
Maldacena (HM) considered a special limit $(J\rightarrow \infty$,
$\lambda$ = fixed, $p$ = fixed, $E-J$ = fixed), with $J$ being the
angular momentum, $E$ the total energy, $p$ the momentum carried
by the string in the spin chain and $\lambda$  the `t Hooft
coupling constant. In this limit, the problem of determining the
spectrum on both (i.e. string and gauge theory) sides becomes
simple. The spectrum consists of an elementary excitation known as
magnon which propagates with a conserved momentum $p$ along the
spin chain. Further, a general class of rigidly rotating string
solutions in $AdS_5\times S^5$ is the spiky string which describes
the higher twist operators from the dual field theory viewpoint.
giant magnons can be visualised as a special limit of such spiky
strings with shorter wavelengths. A large class of such spiky
strings in various asympotically AdS and non-AdS  backgrounds have
been studied in the literature
\cite{Kruczenski:2004wg},\cite{Frolov:2003qc},\cite{Ishizeki:2007we},\cite{Ishizeki:2008tx},\cite{Biswas:2012wu},\cite{Banerjee:2014gga},\cite{Banerjee:2015nha}.

\noindent Therefore, it is interesting to look at the geometric
properties of such string configurations from the world sheet view
point and study normal perturbations (linearized) around the
classical solutions. In this connection, for the case of rotating
superstrings there has been studies \cite{Frolov:2002av} on
computation of leading quantum corrections to the energy spectrum
by expanding the supersymmetric action to quadratic order in
fluctuations near the classical solution. In flat spacetime, where
the string action is Gaussian in the conformal gauge, this would
account for the full string spectrum of states no matter what
classical solutions one starts with.

\noindent In this article, we investigate the classical stability
of the spiky strings using the well-known Jacobi equations
\cite{garriga: 1993},\cite{guven: 1993},\cite{frolov: 1994},
\cite{capovilla: 1995} which govern normal deformations of an
embedded surface. Recent work on such perturbations can be found
in \cite{forini:2015}. We choose to first work with the simplest
spiky strings--the Kruczenski solution in flat spacetime in $2+1$
dimensions and generalize it further to $3+1$ dimensions.  The
rest of the paper is organised as follows. In section 2, we
briefly summarize the solutions and the two different embeddings
of the worldsheet that will be used in our analysis of
perturbations. Section 3 is devoted to study of the perturbation
equations, their solutions and the perturbed profiles, with
illustrative examples. The straightforward extension to a flat
background in $3+1$ dimensions is discussed in Section 4.
Concluding remarks appear in Section 5.

\section{Spiky strings in $2+1$ dimensions:
the Kruczenski and Jevicki-Jin embeddings}

\noindent The Nambu-Goto action for the bosonic string embedded in an
arbitrary $N$ dimensional curved background spacetime with metric
functions $g_{ij}(x)$, is given as
\begin{equation}
S = -T \int d\tau \,\, d\sigma \sqrt{-\gamma} = - T \int  d\tau \,\, 
d\sigma \sqrt{ (\dot{x}\cdot x')^2-\dot{x}^2 {x'}^2} \ .
\end{equation}
where $\gamma$ is the determinant of the induced metric $\gamma_{ab}=g_{ij}
\partial_a x^i \partial_b x^j$ ($a,b=\sigma,\tau$); \, 
$x^i(\tau,\sigma)$ the embedding functions and
$T$ the string tension. We have denoted $\dot x = \partial_{\tau} x$,\,
$ x' = \partial_{\sigma} 
x$, \, $(\dot x \cdot x') = g_{ij}{\dot x}^{i} \, {x'}^{j}$,\, 
${\dot x}^2=g_{ij}{\dot x}^i \,{\dot x}^j$ and
${x'}^2=g_{ij}{ x'}^i \,{x'}^j$.

\noindent Choosing a conformal gauge in which the induced metric is diagonal, 
we have
\begin{equation}
g_{ij}({\dot x}^{i}{\dot x}^{j}+{x'}^{i}{x'}^{j}) = 0, \hspace{0.3in} g_{ij}{\dot x}^{i}{x'}^{j} = 0 \ .
\label{ee2}
\end{equation}

\noindent In this conformal gauge, the string equations of motion 
obtained by varying the action w.r.t. $x^\mu$, take the form:
\begin{equation}
{\ddot x}^{i} - {x^{i ''}} + \Gamma^{i}_{jk} \left
( {\dot x}^{j} {\dot x}^{k} - {x}^{j \prime} {x}^{k\prime}
\right )  = 0 \ .
\label{ee1}
\end{equation}

\noindent We first consider a $2+1$ dimensional flat background spacetime 
with a line element (in $t,\rho, \theta$ coordinates) given as:
\begin{equation}
ds^2 =-dt^2 + d\rho^2 + \rho^2 d\theta^2 \ .
\end{equation}
There are various choices for embeddings which may be used to obtain
string worldsheets which satisfy the string equations of motion (\ref{ee1}) and
the Virasoro constraints (\ref{ee2}). Denoting $\tau$ and $\sigma$ as 
worldsheet coordinates we may choose the following embedding due to 
Jevicki and Jin \cite{jevicki:2008}:
\begin{equation}
t= \tau + f(\sigma) , \hspace{0.1in} \rho=
\rho(\sigma) , \hspace{0.1in} \theta= \omega \tau +
g(\sigma) \ .
\end{equation}
Using the above ansatz in the string equations of motion and
constraints, we arrive at the following equations of motion
\begin{eqnarray}
f''=0,\hspace{0.2in}
\rho''+\rho \left (\omega^2-{g'}^2\right )=0 \\
g'' +2\frac{\rho'}{\rho} g' =0
\end{eqnarray}
and constraints
\begin{eqnarray}
-(1+f'^2) +{\rho'}^2 +\rho^2 \left (\omega^2+{g'}^2 \right )=0
 ,\hspace{0.2in}
f'=\rho^2\omega g'.
\end{eqnarray}
The first integrals of the above equations are:
\begin{eqnarray}
f'=\bar a ,\hspace{0.2in} g'=\frac{\bar a}{\omega \rho^2} \\
{\rho'}^2 = \left (1-\rho^2\omega^2\right )\left (1-\frac{{\bar a}^2}{\omega^2
\rho^2}\right )
\end{eqnarray}
A further integration finally yields the functions $\rho(\sigma)$, $f(\sigma)$
and $g(\sigma)$ and we get
\begin{eqnarray}
t = \tau + {\bar a}\left (\sigma-\sigma_0 \right ), \>\>\> \rho =
\frac{1}{\omega} \left [{\bar a}^2 + (1-{\bar a}^2)\sin^2
\omega(\sigma-\sigma_0) \right ]^{\frac{1}{2}}, \nonumber  \\
\theta = \omega \tau +  \tan^{-1} \left [\frac{1}{\bar a} \tan
\{\omega
(\sigma-\sigma_0)\} \right ] \  \nonumber \\
\end{eqnarray}
where $0<\bar a<1$  and $\sigma_0$ are constants. 

\noindent Let us now convert this solution to Cartesian coordinates which will
enable us to plot the profiles of the unperturbed and perturbed
string worldsheets at fixed values of $\tau$. To do this we need
to use certain coordinate transformations and also some
definitions. We will work with $\sigma_0=0$. The  $\omega
\sigma_0=\frac{\pi}{2}$ case can also be worked out easily. We
define $\rho_0$ and $\rho_1$ as follows:
\begin{equation}
\rho_1 = \frac{1}{\omega} , \hspace{0.1in} \rho_0 =
\frac{\bar a}{\omega} \ ,
\end{equation}
where $\rho_0$ and $\rho_1$ are the minimum and maximum values of
the function $\rho(\sigma)$. Spikes occur at $\rho_1$ whereas
$\rho_0$ correspond to valleys. To go over to Cartesian
coordinates we write $x=\rho \cos \theta$, $y=\rho \sin \theta$
and then use new worldsheet coordinates ($\tau'$ and $\sigma'$)
via a coordinate transformation. As noted by Kruczenski
\cite{Kruczenski:2004wg} the coordinate transformation is given as
\begin{equation}
\tau = -\frac{\rho_1^2}{\rho_0-\rho_1} \tau' - \frac{\rho_0
\rho_1}{\rho_0-\rho_1} \sigma' , \hspace{0.2in}
\sigma = \frac{\rho_1 \rho_0}{\rho_0-\rho_1} \tau' + \frac{
\rho_1^2} {\rho_0-\rho_1}\sigma' \ . \label{eq1}
\end{equation}
Further we define
\begin{equation}
\frac{\rho_1 - \rho_0}{\rho_1 +\rho_0} = \frac{1}{n-1} \ ,
\end{equation}
and note that
\begin{equation}
\bar a = \frac{n-2}{n}, \hspace{0.2in}  \omega =
\frac{1}{\rho_1} \ . \label{eq3}
\end{equation}
If we rewrite $\rho$ and $\theta$ using the above definitions we
see that there are $n$ spikes over the range $0$ to $2\pi$ of
$\sigma'$. The coordinate transformations and the definitions
given above eventually yield the Cartesian embedding functions as
functions of $\tau'$ and $\sigma'$, given as
\begin{eqnarray}
x = \frac{\rho_0}{n-2} \left [ (n-1) \cos \left (\tau'-\sigma'\right ) -
\cos \{(n-1)\left (\tau'+\sigma'\right )\} \right ] \ , \nonumber \\
y = \frac{\rho_0}{n-2} \left [ (n-1) \sin \left
(\tau'-\sigma'\right ) - \sin \{(n-1)\left (\tau'+\sigma' \right
)\} \right ] \ ,
\end{eqnarray}
with $t$ proportional to $\tau'$. We will make use of both the
embeddings in order to understand the perturbations. As mentioned
before, if we use $\omega \sigma_0 =\frac{\pi}{2}$ we will get
back, in Cartesian coordinates, the exact forms given by
Kruczenski \cite{Kruczenski:2004wg}.

\section{Perturbations of spiky strings in $2+1$ dimensions}
Before we embark on writing down the perturbation equations for
our specific solutions, we review the main results on the
perturbations of extremal worldsheets.
\subsection{Perturbation equations for extremal worldsheets}
Assuming (as before) $x^i(\tau,\sigma)$ as the embedding functions and
$g_{ij}$ as the background metric, the tangent vectors to the
worldsheet are defined as
\begin{equation}
e^i_{\tau} = \partial_\tau x^i , \hspace{0.2in}
e^i_{\sigma} = \partial_\sigma x^i ,
\end{equation}
The induced line element is
\begin{equation}
\gamma_{ab} = g_{ij} e^i_a e^j_b \ ,
\end{equation}
where the $a,b...$ denote worldsheet indices (here $\tau$,
$\sigma$). The normals to the worldsheet $n^i_{(\alpha)}$ satisfy
the following relations
\begin{equation}
g_{ij} n^i_{(\alpha)}n^{j}_{(\beta)} = \delta_{\alpha\beta}
 ,\hspace{0.2in} g_{ij}n^i_{(\alpha)} e^j_{a}=0 \ ,
\end{equation}
where $\alpha=1..,N-2$ where $N$ is the dimension of the
background spacetime. The last condition is valid for all $\alpha$
and $a$. The extrinsic curvature tensors $K_{ab}^{(\alpha)}$ along
each normal $n^i_{(\alpha)}$ of the embedded worldsheet are defined
as
\begin{equation}
K_{ab}^{(\alpha)} = -g_{ij}( e^k_{a}\nabla_k e^i_b) n^{j(\alpha)}
\ .
\end{equation}
One can check that the equations of motion correspond to the
condition that $K^{(\alpha)} = \gamma^{ab} K_{ab}^{(\alpha)} =0$
which means that the worldsheet is an extremal surface with zero
value for the trace of each extrinsic curvature tensor. Normal
perturbations are defined using a set of scalar fields
$\phi^{(\alpha)}$ along each normal. We have
\begin{equation}
\delta x^i = \phi^{(\alpha)} n^i_{(\alpha)} \ ,
\end{equation}
as the resultant perturbation of the worldsheet (i.e. $x^i
\rightarrow x^i + \delta x^i$). For an extremal worldsheet
($\gamma^{ab}K_{ab}^{(\alpha)}=0$) satisfying the equations of
motion and the Virasoro constraints, the scalars
$\phi^i_{(\alpha)}$ satisfy the following equations (Jacobi
equations) which follow from the second variation of the
Nambu-Goto action.
\begin{equation}
\frac{1}{\Omega^2} \left ( - \frac{\partial^2}{\partial \tau^2} +
\frac{\partial^2}{\partial \sigma^2} \right ) \phi^{(\alpha)} +
\left (M^2\right )^{(\alpha)}_{(\beta)} \phi^{(\beta)} = 0 \ ,
\end{equation}
where
\begin{equation}
\left (M^2\right )^{(\alpha)}_{(\beta)} = K_{ab}^{(\alpha)}K^{ab}_
{(\beta)} +R_{ijkl} e^j_a e^{l\,\,a} n^{i\,\,(\alpha)}
n^k_{(\beta)} \ ,
\end{equation}
and $\Omega^2(\tau,\sigma)$ is the conformal factor of the
conformally flat worldsheet line element. Solving this equation
for the perturbation scalars one can learn about the stability of
the extremal worldsheet under perturbations. Note that the Jacobi
equations turn out to be a family of coupled variable `mass'  wave
equations for the perturbation scalars $\phi^{(\alpha)}$. In
general, they are difficult to solve exactly, even for the
simplest cases. However, fortunately, for the cases discussed
below we do find exact solutions quite easily.

\noindent We may also note the fact that, in general, the
worldsheet covariant derivative (which arises in the first term in Eqn. 3.6)
can have a contribution
from the normal fundamental form (extrinsic twist potential)
defined as
$\omega_a^{\,\,\alpha\beta}=
g_{ij}\, \left ( e^k_a\nabla_k \, n^{i\,\alpha}\right )\,n^{j\,\beta}$.
However, for hypersurfaces and for the normals considered in the
$3+1$ dimensional case to be discussed later, the $\omega_a^{\alpha\beta}$
are identically zero.

\subsection{The case of spiky strings}
In order to find the perturbation equations for our case we need
to first write down the tangent, normal, induced metric and
extrinsic curvature for the world sheet. Here, we use the
Jevicki-Jin embedding mentioned earlier. The tangent vector to the
worldsheet is given as:
\begin{equation}
e^{i}_\tau = \left (1,0,\omega\right ) ,
\hspace{0.2in} e^{i}_\sigma = \left ( f', \rho',g'\right ) \ .
\end{equation}
Using the expressions for $f$, $g$ and $\rho$ quoted earlier the
induced line element turns out to be
\begin{equation}
ds^2= \left (1-\rho^2\omega^2\right ) \left (-d\tau^2 +
d\sigma^2\right ) \ .
\end{equation}
The normal to the worldsheet is given as:
\begin{equation}
n^{i} = \left (\tan \omega \sigma, -\frac{\bar a}{\omega \rho},
\frac{\tan \omega \sigma}{\omega \rho^2}\right ) \ .
\end{equation}
The extrinsic curvature tensor turns out to be
\begin{equation}
K_{ab} = \begin{pmatrix} -\bar a \omega & -\omega \cr -\omega &
-\bar a \omega \end{pmatrix} \ .
\end{equation}
Hence the quantity $K_{ab}K^{ab}$ is given as:
\begin{equation}
K_{ab} K^{ab} = -\frac{2\omega^2 (1-\bar
a^2)}{(1-\rho^2\omega^2)^2}=-{}^2 R \ ,
\end{equation}
where ${}^{2}R$ is the worldsheet Ricci scalar. Given that $\bar a<1$,
we note that ${}^{2}R$ is positive and blows up to positive infinity
at the location of the spikes, i.e. wherever $\rho^2\omega^2=1$.

Since the background is flat the Riemann tensor is zero and,
therefore, the equation for the perturbation scalar (here only one
scalar because there is only one normal) turns out to be:
\begin{equation}
\left ( -\frac{\partial^2}{\partial \tau^2} +
\frac{\partial^2}{\partial \sigma^2} \right )\phi -
\frac{2\omega^2 (1-\bar a^2)}{(1-\rho^2\omega^2)^2} \phi =0 \ .
\end{equation}
Using the functional form of $\rho(\sigma)$ one can reduce this equation to
\begin{equation}
\left (- \frac{\partial^2}{\partial \tau^2} +
\frac{\partial^2}{\partial \sigma^2} \right ) \phi - 2\omega^2
\sec^2\omega \sigma \phi =0 \ .
\end{equation}
Let us take a simple separable ansatz of the form
\begin{equation}
\phi = e^{i\beta \tau} P(\sigma) \ .
\end{equation}
We end up with an equation for $P(\sigma)$ given as
\begin{equation}
\frac{d^2P}{d\sigma^2} +\left (\beta^2 - 2 \omega^2 \sec^2
\omega\sigma\right ) P =0 \ .
\end{equation}
A simple transformation $\xi = \omega\sigma$ reduces this equation to
\begin{equation}
\frac{d^2P}{d\xi^2} +\left (\gamma^2 - 2 \sec^2 \xi\right )
P =0
\end{equation}
where $\beta = \gamma \omega$. It turns out that this is an
equation which we encounter in quantum mechanical potential
problems--a special case of the well-known Poschl-Teller potential
\cite{flugge:book}. One can surely solve this equation exactly but
before we do that let us try and look at a simple and obvious
solution. For $\gamma=2$ we have a solution for $P(\xi)$ given as
\begin{equation}
P(\xi) = \epsilon \rho_0 \cos^2 \xi \ , \label{s01}
\end{equation}
where $\epsilon$ is a constant and we may relate it to the
amplitude of the perturbation. It must be emphasized that $\epsilon$
has to be small in value (i.e. $\epsilon<<1$) in order that the
deformation is genuinely a perturbation. This also follows from the
relative magnitudes of the two terms in the expression for $\rho'$ below. 
Since the minimum value of $\rho$ is $\rho_0$ and the perturbation
goes as $\epsilon\rho_0$, one ie required to have $\epsilon<<1$. 
More generally, the Jacobi equations are not even valid equations 
for the perturbations when the perturbations themselves become large.

\noindent Let us now write down the
perturbations for this mode (the abovestated $\gamma=2$ solution $P(\xi)$). 
For the $\tau$ part of the full solution
we take its real part (i.e. $\cos \beta \tau = \cos 2 \omega
\tau$). One may also work with the imaginary part. We find
\begin{eqnarray}
t'= t+ \phi n^0 = t + \epsilon \rho_0\cos 2\omega \tau \sin \omega \sigma
\cos \omega \sigma, \nonumber \\
\rho' = \rho + n^1 \phi = \rho - \epsilon \rho_0\frac{\bar a \cos 2 \omega \tau \cos^2 \omega \sigma}{\sqrt{{\bar a}^2
+(1-{\bar a}^2) \sin^2 \omega \sigma}}, \nonumber \\
\theta'= \theta + n^2 \phi = \theta + \epsilon \rho_0 \frac{\omega
\cos 2\omega \tau \sin \omega \sigma \cos \omega \sigma}{{\bar
a}^2 +(1-{\bar a}^2) \sin^2 \omega \sigma} \ . \nonumber \\
\end{eqnarray}
Let us now try to understand how this perturbation affects the
string profile. We need to switch back to Cartesian coordinates to
get a clearer picture here. Defining $x' = \rho' \cos\theta'$ and
$y'=\rho' \sin \theta'$ we use $\rho' = \rho +\delta \rho$ and
$\theta' = \theta +\delta\theta$ to eventually obtain
\begin{eqnarray}
t'= t+\delta t,\>\>\>  x' = x + \frac{\delta \rho}{\rho} \,\,x -
\delta \theta\,\, y \ , \>\>\>  y' = y +
\frac{\delta\rho}{\rho}\,\, y + \delta \theta \,\,x \ .  \label{eqapx}
\end{eqnarray}
In the above expressions, we will now substitute $\delta t$,
$\frac{\delta \rho}{\rho}$, $\delta \theta$ and switch back to the
$(\tau', \sigma')$ coordinates on the worldsheet. Thereafter,
taking small values of $\epsilon$ we will obtain the perturbed
worldsheet and compare it with the unperturbed worldsheet. This
will help us get an idea of what the normal perturbations do to
the spiky string.

\noindent Let us begin with the solutions for $P(\xi)$. For $\gamma=2$ we
have noted above that $P(\xi) =\epsilon \rho_0\cos^2 \xi$ is a
solution. The other linearly independent solution has an overall
$\sec \xi$ and is divergent. This feature persists for all pairs
of linearly independent solutions for different values of
$\gamma$. We have one oscillatory solution and another divergent
solution. We can write the general solutions using hypergeometric
functions (see details later). We now move on to writing down
$\delta t$, $\frac{\delta \rho}{\rho}$ and $\delta \theta$ using
the $(\tau',\sigma')$ coordinates for the $P(\xi) = \epsilon
\rho_0 \cos^2 \xi$ solution with $\gamma =2$. This gives
\begin{eqnarray}
\delta t = \frac{\epsilon \rho_0}{4} \left [ \sin 2(\tau'-\sigma') -\sin 2(n-1)(\tau'-\sigma')\right ] \\
\frac{\delta \rho}{\rho} =
- \epsilon \frac{(n-2)^2 \cos \{n \tau'+(n-2)\sigma'\}
\cos^2 \{\frac{n-2}{2}\tau' +\frac{n}{2} \sigma'\}}{(n-2)^2 +4(n-1)
\sin^2 \{\frac{n-2}{2}\tau' + \frac{n}{2}\sigma'\}} \\
\delta \theta = -
\epsilon \frac{n(n-2) \cos \{n \tau'+(n-2)\sigma'\}
\cos \{\frac{n-2}{2}\tau' +\frac{n}{2} \sigma'\}
\sin \{\frac{n-2}{2}\tau' +\frac{n}{2} \sigma'\}}
 {(n-2)^2 +4(n-1)
\sin^2 \{\frac{n-2}{2}\tau' + \frac{n}{2}\sigma'\}} \ .
\end{eqnarray}
These expressions can now be substituted in those for $t'$, $x'$,
$y'$ in order to obtain the perturbed worldsheet. The simplest
case is with $n=3$. Here, we have $t'$, $x'$ and $y'$ as follows.
\begin{eqnarray}
t'= 4\tau' + \frac{\epsilon \rho_0}{4} \left [ \sin[2(\tau'-\sigma')] -
\sin[4(\tau'+\sigma')]\right ] \ , \\
x' =  A(\tau',\sigma') \,\, x + B(\tau',\sigma') \,\,y  \ , \\
y'= -B(\tau',\sigma') \,\, x + A(\tau',\sigma') \,\, y \ ,
\end{eqnarray}
where
\begin{eqnarray}
A(\tau',\sigma') = \left ( 1-\epsilon \frac{\cos (\sigma'+3\tau') \cos^2 (\frac{3\sigma'}{2}+\frac{\tau'}{2})}{1+8\sin^2 (\frac{3\sigma'}{2}+\frac{\tau'}{2})}\right ) \\
B(\tau',\sigma') = \left (\epsilon \frac{3 \cos (\sigma'+ 3\tau')\sin (\frac{3\sigma'}{2}+\frac{\tau'}{2})
\cos (\frac{3\sigma'}{2} +\frac{\tau'}{2})}{1+ 8 \sin^2(\frac{3\sigma'}{2}+\frac{\tau'}{2})}\right )
\end{eqnarray}
and $x$ and $y$ are given as
\begin{eqnarray}
x = 2\cos (\tau'-\sigma') - \cos 2 (\tau'+\sigma') \ , \>\>\>  y=
2\sin (\tau'-\sigma') -\sin 2 (\tau'+\sigma') \ . \nonumber \\
\end{eqnarray}
Note that, in writing down the above an overall factor of $\rho_0$
has been scaled out in the definition of the coordinates.

\begin{center}
\begin{figure}[h]
 \begin{minipage}{14pc}
 \includegraphics[width=11pc]{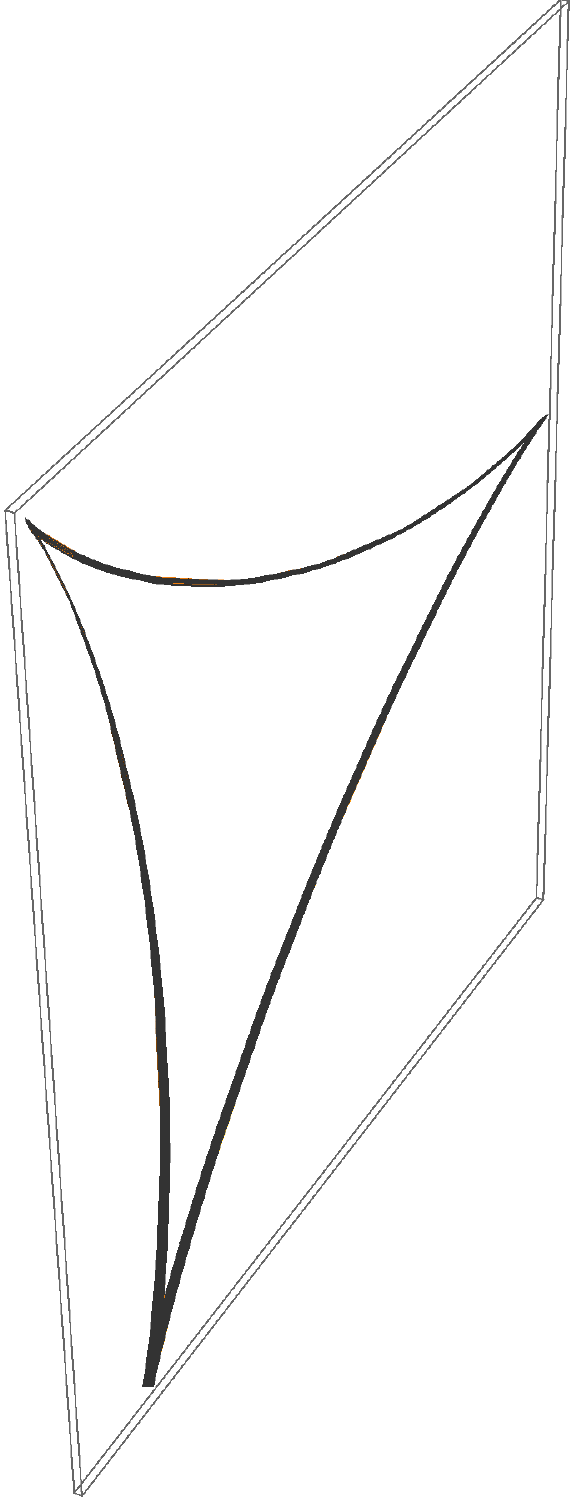}
 \caption{$n=3$, $\tau'\equiv (4.0,4.01)$, unperturbed.}
 \end{minipage}
\hspace{1in}
 \begin{minipage}{14pc}
 \includegraphics[width=11pc]{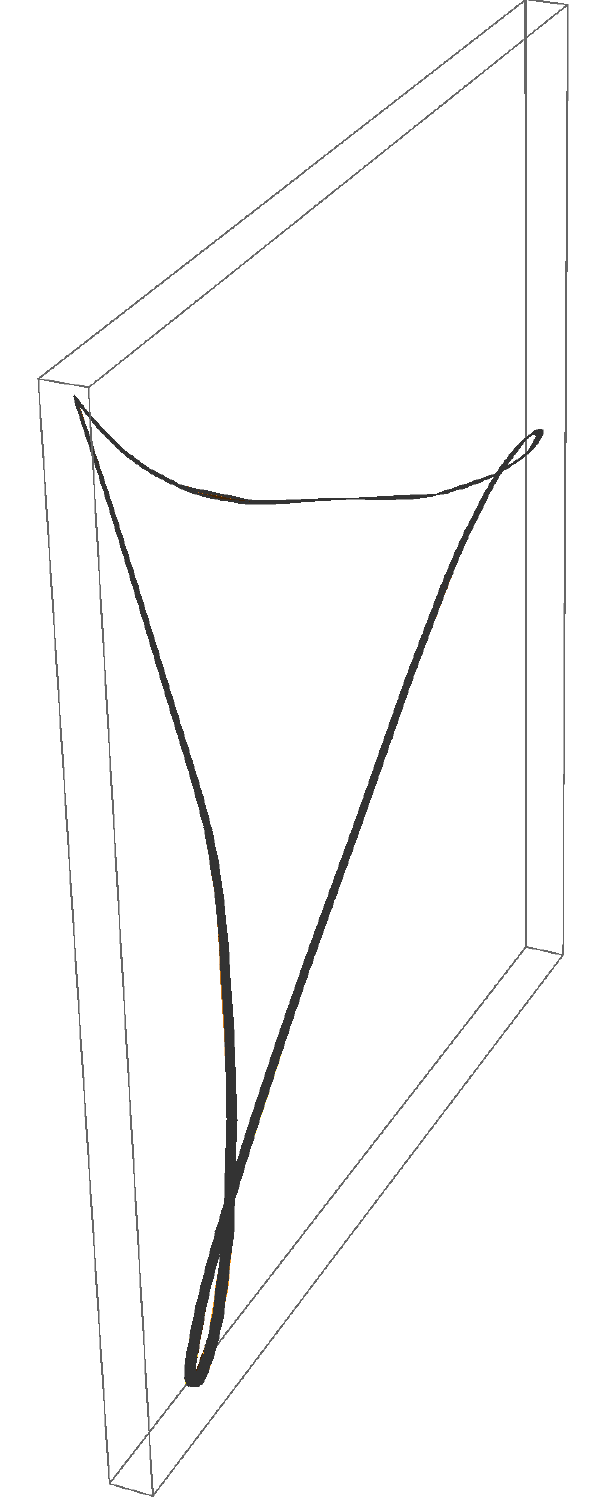}
 \caption{$n=3$, $\epsilon=0.2$, $\tau'\equiv (4.0,4.01)$, perturbed.}
 \end{minipage}
 \end{figure}
\end{center}

\noindent Figures 1-6 demonstrate the scenario for $n=3$. Figure 1
shows the worldsheet over a small range of $\tau'$ (i.e. $\tau'=4.0$ to
$\tau'=4.01$). One can see the three spikes. The very slight
rigid rotation of the profile along $\tau'$
is visible through the thickened character
of the black lines. Figure 2 represents the perturbed worldsheet
profile for the same range of $\tau'$. One notices that the
worldsheet is now spread across the $t'$ direction because $t'$ is
dependent on both $\tau'$ and $\sigma'$. In Figure 2, the spikes
are rounded off and the string spreads out-- there are no
self-intersections. Figure 3 shows the worldsheet profile near the
spikes whereas Figure 4 shows how a spike location gets rounded
off due to the perturbation. Thus, the effect of the perturbation seems to be in
rounding off the spikes. Similar observations were made in the
presence of large angular momentum in
\cite{Ishizeki:2008tx}, \cite{mosaffa:2007}.

 \begin{figure}[h]
 \begin{minipage}{14pc}
 \includegraphics[width=11pc]{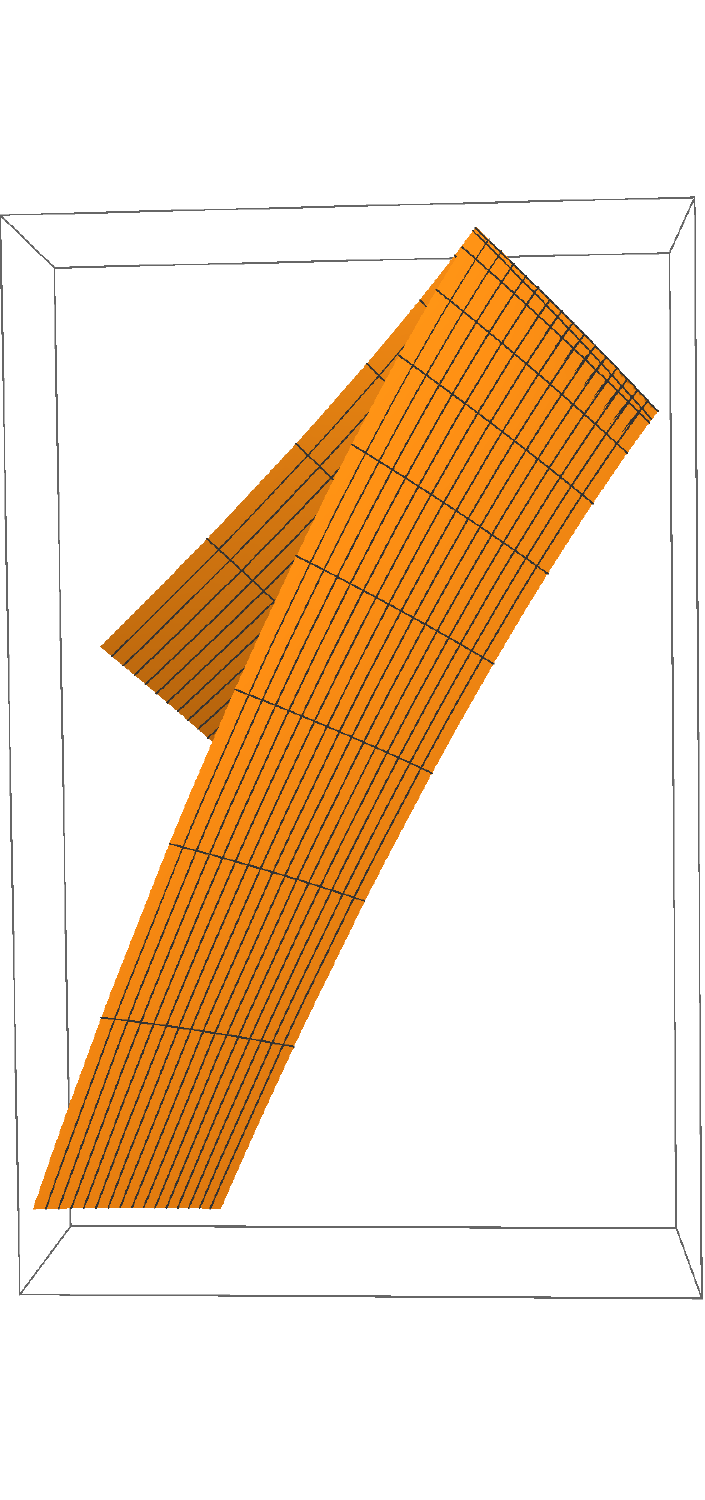}
 \caption{$n=3$, unperturbed, $\tau'\equiv(4.0,4.05)$,
$\sigma'\equiv (0.4 \pi,0.7 \pi)$.}
 \end{minipage}
\hspace{1in}
 \begin{minipage}{14pc}
 \includegraphics[width=11pc]{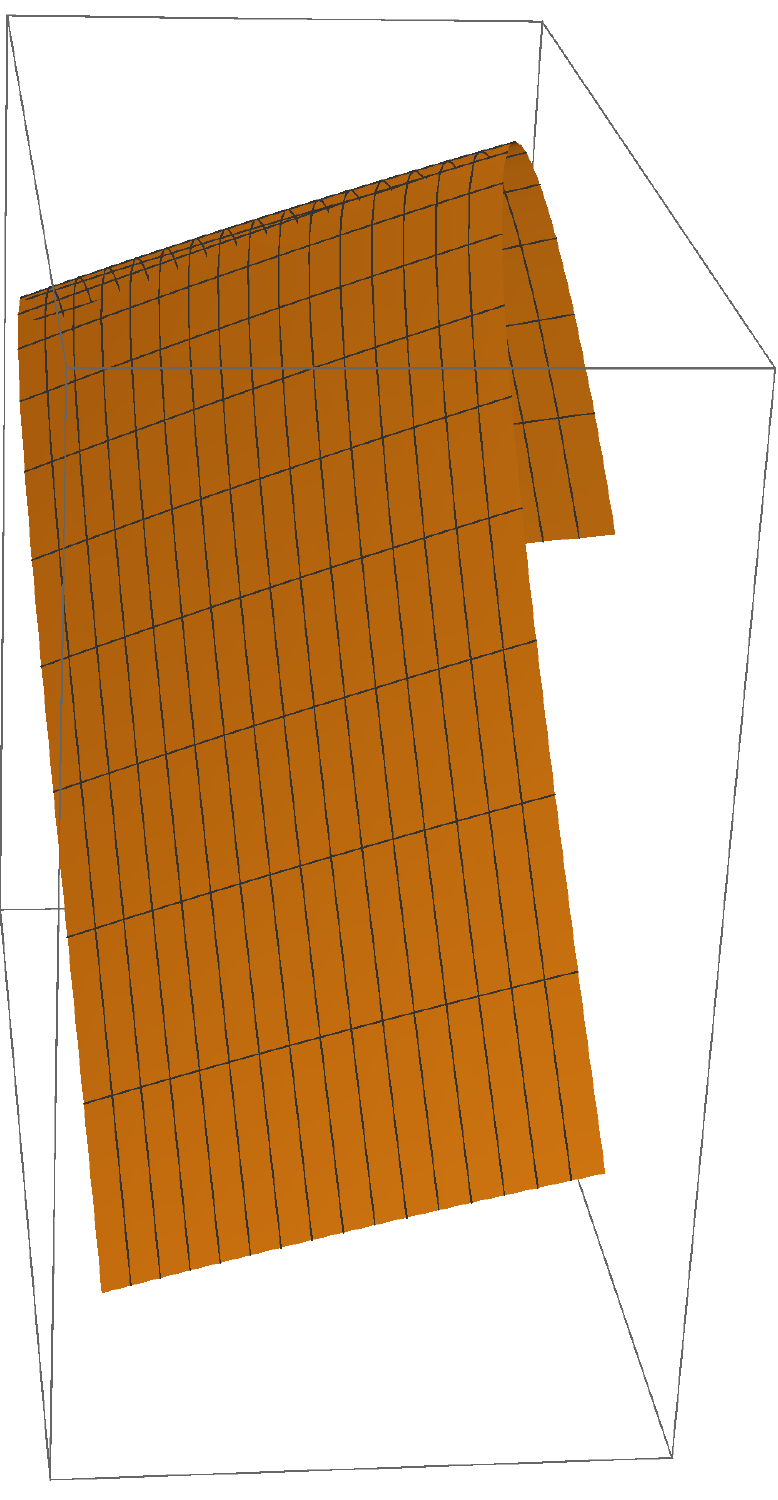}
 \caption{$n=3$, perturbed, $\epsilon=0.24$, $\tau'\equiv(4.0,4.05)$, $\sigma'\equiv (0.5 \pi,0.7 \pi)$. }
 \end{minipage}
 \end{figure}

\noindent It is easy to repeat the above calculation to obtain the other
modes. Here too we have found similar features. We do not discuss
them in any further detail.

\noindent The general solutions of the
second order differential equation for $P$ can be
written as a superposition of the two linearly independent ones, both of
which involve hypergeometric functions. Note that the differential
equation is a special case of the well-known Poschl-Teller potential
problem in quantum mechanics \cite{flugge:book}.
We can write
\begin{equation}
P(\xi)= C_1 P_1 + C_2 P_2 \ ,
\end{equation}
where the $P_1$ and $P_2$ are given by
\begin{equation}
P_1 =\epsilon \cos^2{\xi}\,\,{}_2 F_1(\frac{2+\gamma}{2},
\frac{2-\gamma}{2}, \frac{1}{2};\sin^2{\xi}) \ , \label{s1}
\end{equation}

\vspace{0.3in}
\begin{equation}
P_2 = \epsilon \sin{\xi} \cos^2{\xi} \,\,{}_2 F_1(\frac{3+\gamma}{2}, \frac{3-\gamma}{2}, \frac{3}{2}; \sin^2{\xi})
\label{s2}
\end{equation}
where
\begin{equation}
\gamma = 2+2\nu,\hspace{0.2in}\nu = 0,1,2..... \ .
\end{equation}
Each allowed value of $\nu$ yields the corresponding fluctuation
mode of a given configuration. As before, we include here the
$\epsilon$, which is the amplitude of the perturbation. For every
allowed value of $\gamma$, $P_1$ represents an oscillatory
solution whereas $P_2$ is a divergent solution.

\begin{figure}[h]
\begin{minipage}{16pc}
\includegraphics[width=16pc]{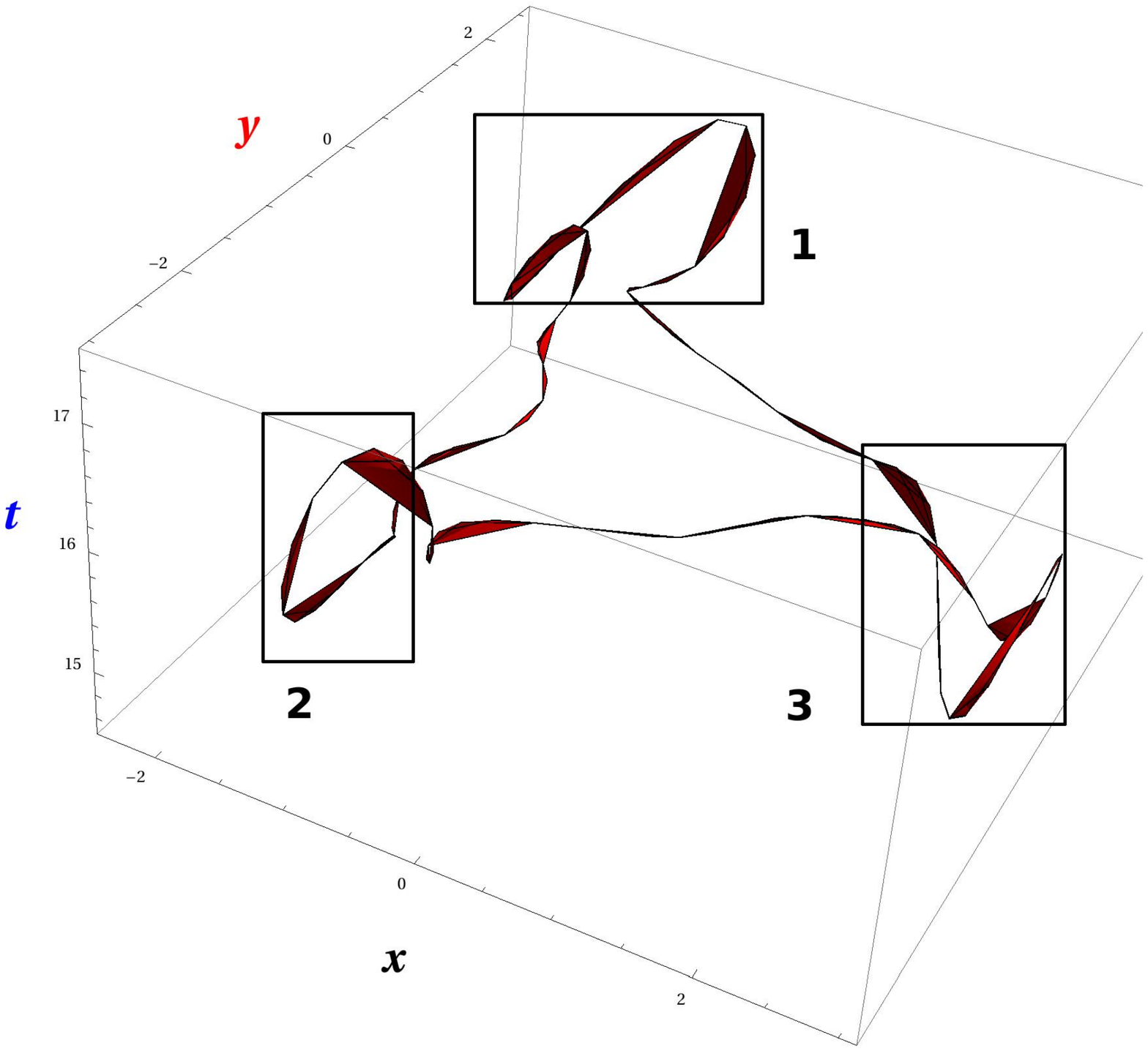}
\caption{$n=3$ , $\gamma = 4$ , $\tau'\equiv (4 , 4.01)$, $\epsilon = 0.1$, $\sigma' \equiv (0 , 2\pi)$.}
\label{g1}
 \end{minipage}
\hspace{0.5in}
\begin{minipage}{16pc}
 \includegraphics[width=16pc]{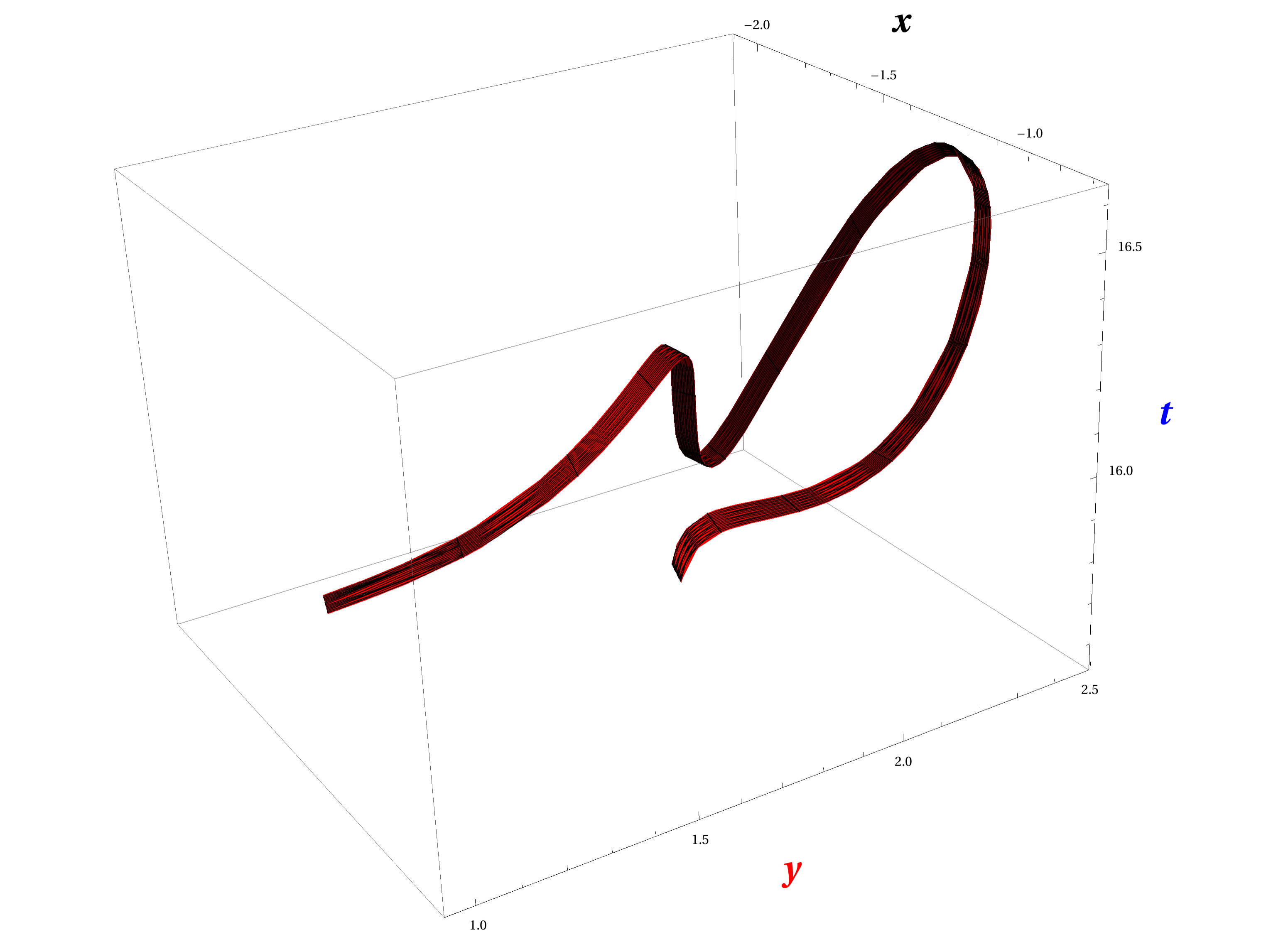}
 \caption{{\bf Region $1$ in Fig. 5}, $n=3$ , $\gamma = 4$ , $\tau'\equiv (4 , 4.01)$, $\epsilon = 0.1$,$\sigma' \equiv (\frac{\pi}{3},0.8\pi)$.}
\label{g11}
 \end{minipage}
\begin{minipage}{16pc}
 \includegraphics[width=16pc]{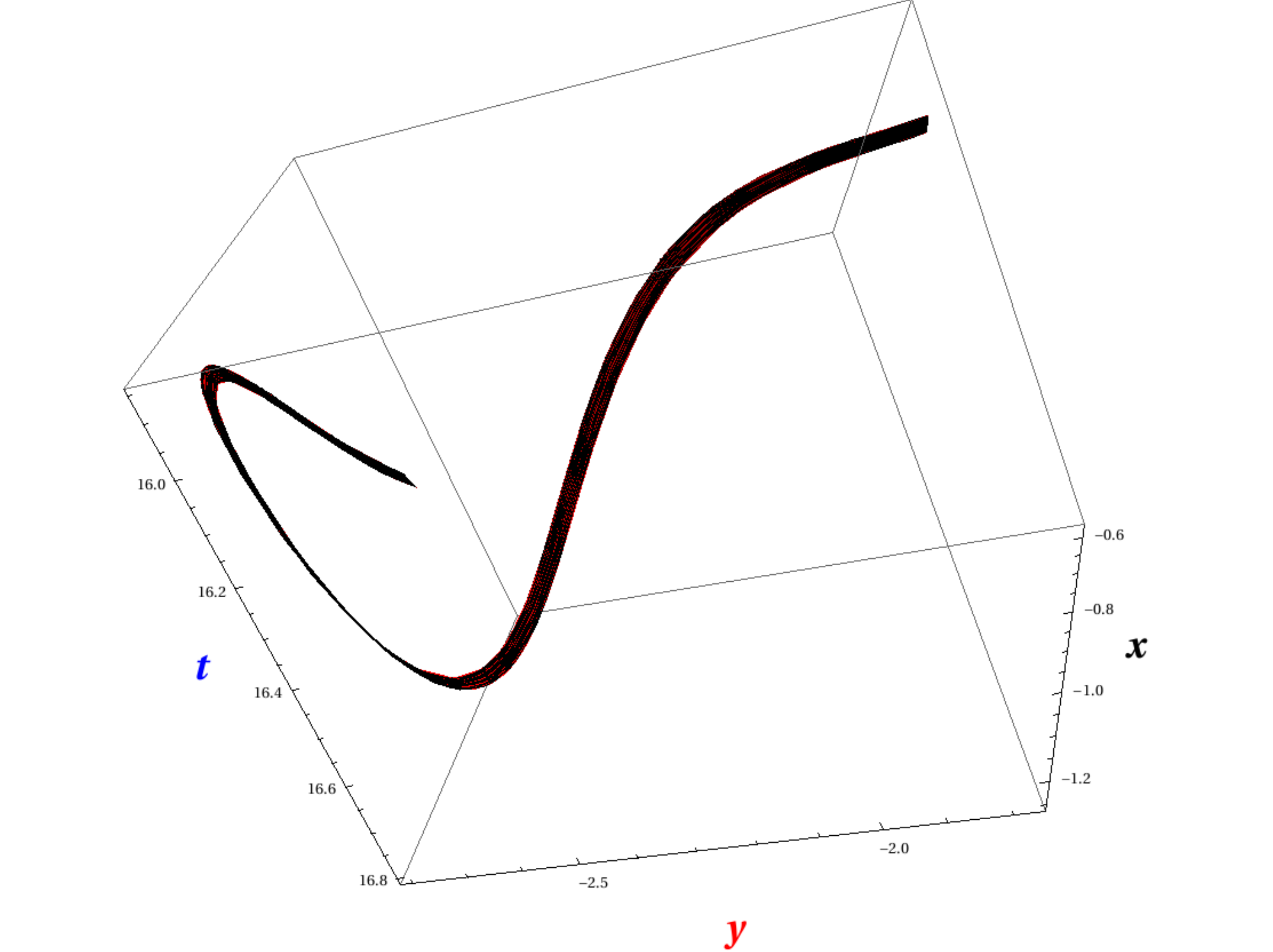}
 \caption{{\bf Region $2$ in Fig. 5}, $n=3$ , $\gamma = 4$ , $\tau'\equiv (4 , 4.01)$, $\epsilon = 0.1$, $\sigma' \equiv (1.7\pi,2\pi)$.}
\label{g14}
 \end{minipage}
\hspace{0.7in}
\begin{minipage}{16pc}
 \includegraphics[width=16pc]{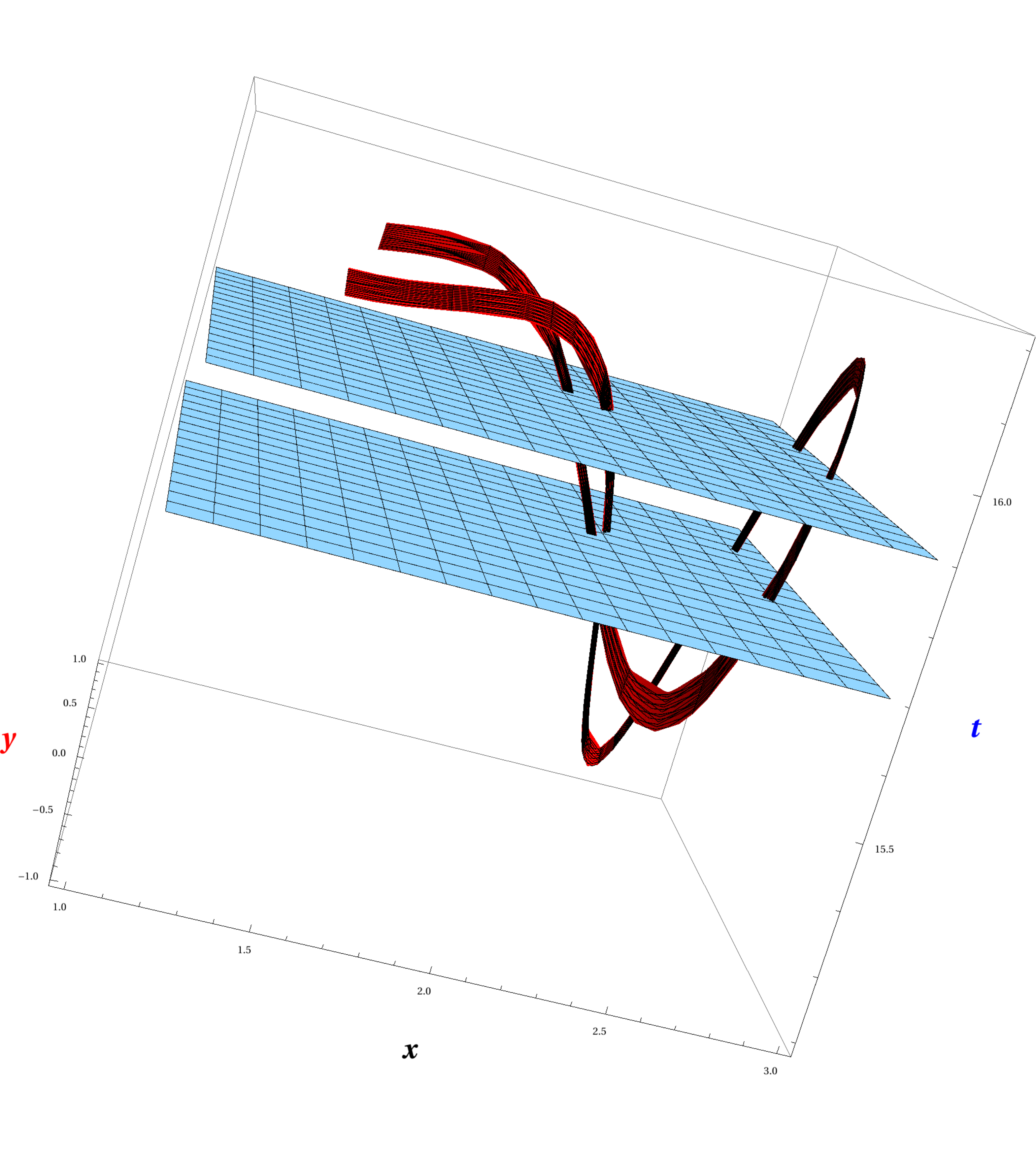}
 \caption{{\bf Region $3$ in Fig. 5}, $n=3$ , $\gamma = 4$ , $\tau'\equiv (4 , 4.01)$, $\epsilon = 0.1$, $\sigma' \equiv (\pi,\frac{3\pi}{2})$.}\label{g12}
 \end{minipage}
\end{figure}

\noindent Let us now write down the perturbations for $P_1$. We find
\begin{eqnarray}
t'= t+ \phi n^0 = t + \epsilon \rho_0 \cos \gamma \omega \tau \sin \omega \sigma
\cos \omega \sigma \,\,{}_2 F_1(\frac{2+\gamma}{2}, \frac{2-\gamma}{2}, \frac{1}{2};
\sin^2{\omega \sigma}) \ , \\
\rho' = \rho + n^1 \phi = \rho - \epsilon \rho_0 \frac{\bar a \cos \gamma \omega \tau \cos^2 \omega \sigma \,\,
{}_2 F_1(\frac{2+\gamma}{2}, \frac{2-\gamma}{2}, \frac{1}{2}; \sin^2{\omega \sigma})}
{\sqrt{{\bar a}^2
+(1-{\bar a}^2) \sin^2 \omega \sigma}} \,  \\
\theta'= \theta + n^2 \phi = \theta + \epsilon \rho_0 \frac{\omega \cos \gamma \omega \tau \sin \omega \sigma \cos \omega \sigma \,\,
{}_2 F_1(\frac{2+\gamma}{2}, \frac{2-\gamma}{2}, \frac{1}{2};
\sin^2{\omega \sigma})}{{\bar a}^2
+(1-{\bar a}^2) \sin^2 \omega \sigma}
\end{eqnarray}
As before,  defining $x' = \rho' \cos\theta'$ and $y'=\rho' \sin
\theta'$ and with $\rho' = \rho +\delta \rho$ and $\theta' =
\theta +\delta\theta$ we evaluate  $\delta t$, $\frac{\delta
\rho}{\rho}$ and $\delta \theta$ using the $(\tau',\sigma')$
coordinates to finally obtain $t'$, $x'$ and $y'$ which are as
follows
\begin{eqnarray}
t'= t+\frac{\epsilon \rho_0}{4} [\sin[\{(\frac{\gamma}{2}-1)n+2\}\tau'+\{(\frac{\gamma}{2}-1)n-2\}\sigma']-\sin\{(\frac{\gamma}{2}+1)n-2\}(\tau'+\sigma')]
 \nonumber \\ {}_2 F_1(\frac{2+\gamma}{2}, \frac{2-\gamma}{2}, \frac{1}{2}; \sin^2 \frac{(n-2)\tau'+ n\sigma'}{2}) \hspace{0.5in}
\end{eqnarray}
\begin{eqnarray}
x' =  C(\tau',\sigma') \,\, x + D(\tau',\sigma') \,\,y \\
y'= -D(\tau',\sigma') \,\, x + C(\tau',\sigma') \,\, y
\end{eqnarray}
where $C$ and $D$ are
\begin{equation}
C=  1-\frac{\epsilon (n-2)^2 \cos^2{\frac{n\sigma'+(n-2)\tau'}{2}}
 \cos{\frac{\gamma (n-2)\sigma'+\gamma n \tau'}{2}} {}_2 F_1 (\frac{2+\gamma}{2}, \frac{2-\gamma}{2}, \frac{1}{2};
\sin^2{\frac{n \sigma'+(n-2)\tau'}{2}})}{(n-2)^2+4 (n-1)
\sin^2{\frac{n \sigma'+(n-2)\tau'}{2}}} \ ,
\end{equation}
\begin{equation}
D= \frac{\epsilon n (n-2) \sin \left [n \sigma'+(n-2)\tau'\right ]
\cos{\frac{\gamma (n-2) \sigma'+\gamma n\tau'}{2}}\, {}_2 F_1
(\frac{2+\gamma}{2}, \frac{2-\gamma}{2}, \frac{1}{2};
\sin^2{\frac{n \sigma'+(n-2)\tau'}{2}})}{2(n-2)^2 + 8 (n-1)
\sin^2{\frac{n \sigma'+(n-2)\tau'}{2}}} \ .
\end{equation}

\noindent Figures 5 and 6 show the unperturbed and perturbed
profiles for $n=3$ and $\gamma=4$. We can see that, as for
$\gamma=2$, the spikes are rounded off, as is visible in the plot
for a small range of $\sigma'$ (corresponding to region $1$ in Figure 5), 
shown in Figure 6. In Figures 7 and 8 we
plot the worldsheets over other different ranges of $\sigma'$ (corresponding to
the regions 2 and 3 in Figure 5) but
for the same range of $\tau'$.

\subsection{Dual spiky strings and perturbations}

\noindent In this section we discuss the T-dual solutions of the spiky strings
first obtained in \cite{mosaffa:2007}.
As before, we assume a $2+1$ dimensional flat background spacetime in $t,\rho, \theta$ coordinates. To arrive at the dual spikes we
use the  dual Jevicki-Jin embedding given by
\begin{equation}
t= \sigma + f(\tau), \>\>\> \rho= \rho(\tau), \>\>\>  \theta=
\omega \sigma + g(\tau) \ .
\end{equation}
This embedding can be obtained from the original Jevicki-Jin embedding by
exchanging the world sheet coordinates $\tau$ and $\sigma$. Later, we will see
why it is necessary to have the dual embedding in order to obtain the
dual spikes.

\noindent Using the above in the string equations of motion and constraints we
can obtain the functions $\rho(\tau)$, $f(\tau)$ and
$g(\tau)$. The string equations of motion and the Virasoro
constraints are of the same form,
with $\tau$ derivatives now replacing the $\sigma$ derivatives.
Finally, we write the dual solution as,
\begin{eqnarray}
t = \sigma + {\bar a} \tau , \>\>\> \rho = \frac{1}{\omega} \left
[1 + ({\bar a}^2-1)\sin^2 \omega \tau \right ]^{\frac{1}{2}} ,
 \>\>\> \theta = \omega \sigma + \tan^{-1} \left [\bar a \tan \{\omega
\tau\} \right ] \ . \nonumber \\
\end{eqnarray}

\noindent Here $\bar a= \frac{\rho_0}{\rho_1} >1$,
whereas, for the spike solution, $\bar a < 1$. The relation between
$\rho_0$ and $\rho_1$ is
\begin{equation}
\frac{\rho_0 - \rho_1}{\rho_1 +\rho_0} = \frac{1}{n-1} \ ,
\end{equation}
and
\begin{equation}
\bar a = \frac{n-2}{n}, \hspace{0.2in}\omega = \frac{1}{\rho_1} \
. \label{eq4}
\end{equation}
To write the dual spike solution in Cartesian coordinates we will need to
use a coordinate transformation similar to the one mentioned earlier
but with $\tau$ and $\sigma$ exchanged. Working through some
straightforward algebra we arrive at
\begin{eqnarray}
x = \frac{\rho_0}{n} \left [ (n-1) \cos \left (\tau'-\sigma'\right )
+ \cos \{(n-1)\left (\tau'+\sigma'\right )\} \right ] \, \nonumber \\
y = \frac{\rho_0}{n} \left [ (n-1) \sin \left (\tau'-\sigma'\right
) - \sin \{(n-1)\left (\tau'+\sigma' \right )\} \right ] \ .
\end{eqnarray}
It is important to realise that in the Jevicki-Jin conformal gauge
the dual spiky string must be obtained using the dual embedding
which has $\tau$ dependent functions (as opposed to $\sigma$
dependent functions in the spiky string case). The reason behind
this fact may be traced to the correct signature of the induced
metric, which is Lorentzian only when the dual embedding is used.

\begin{figure}[h]
\begin{minipage}{12pc}
\includegraphics[width=10pc]{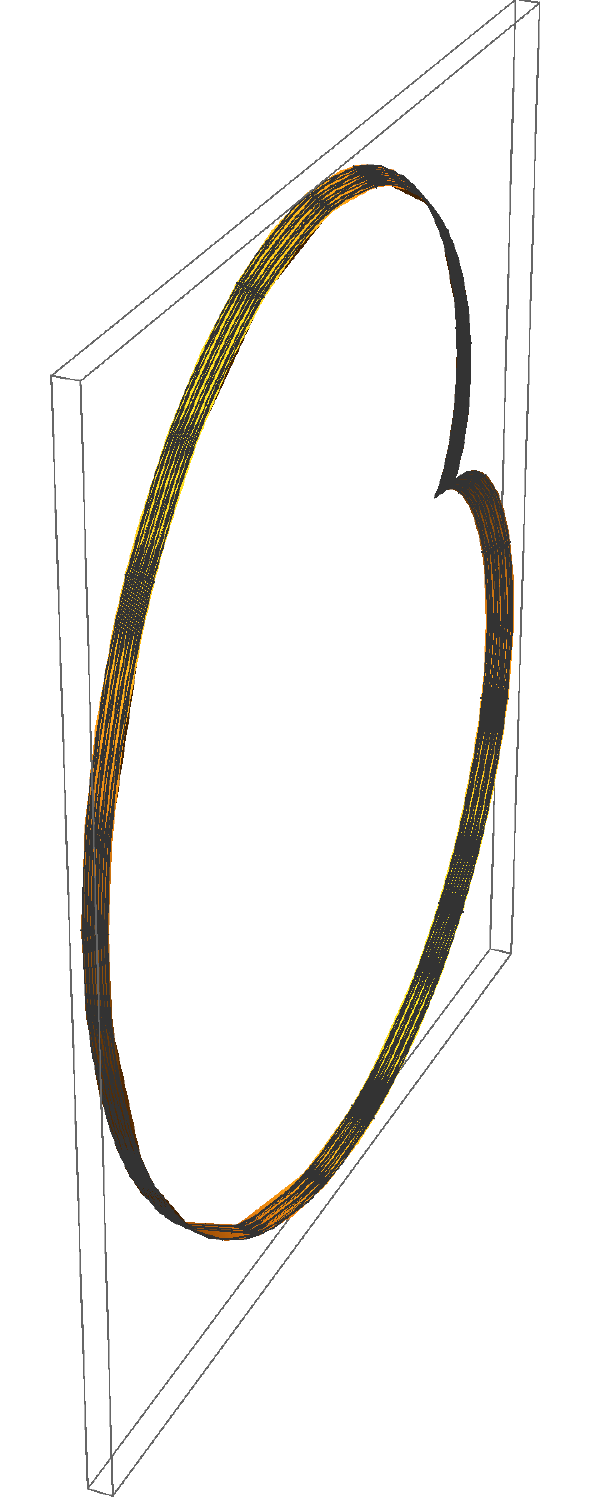}
\caption{$n=3$, $\tau'\equiv (4.0,4.03)$. Unperturbed.}
\label{d31}
\end{minipage}
\hspace{1.2in}
\begin{minipage}{12pc}
\includegraphics[width=10pc]{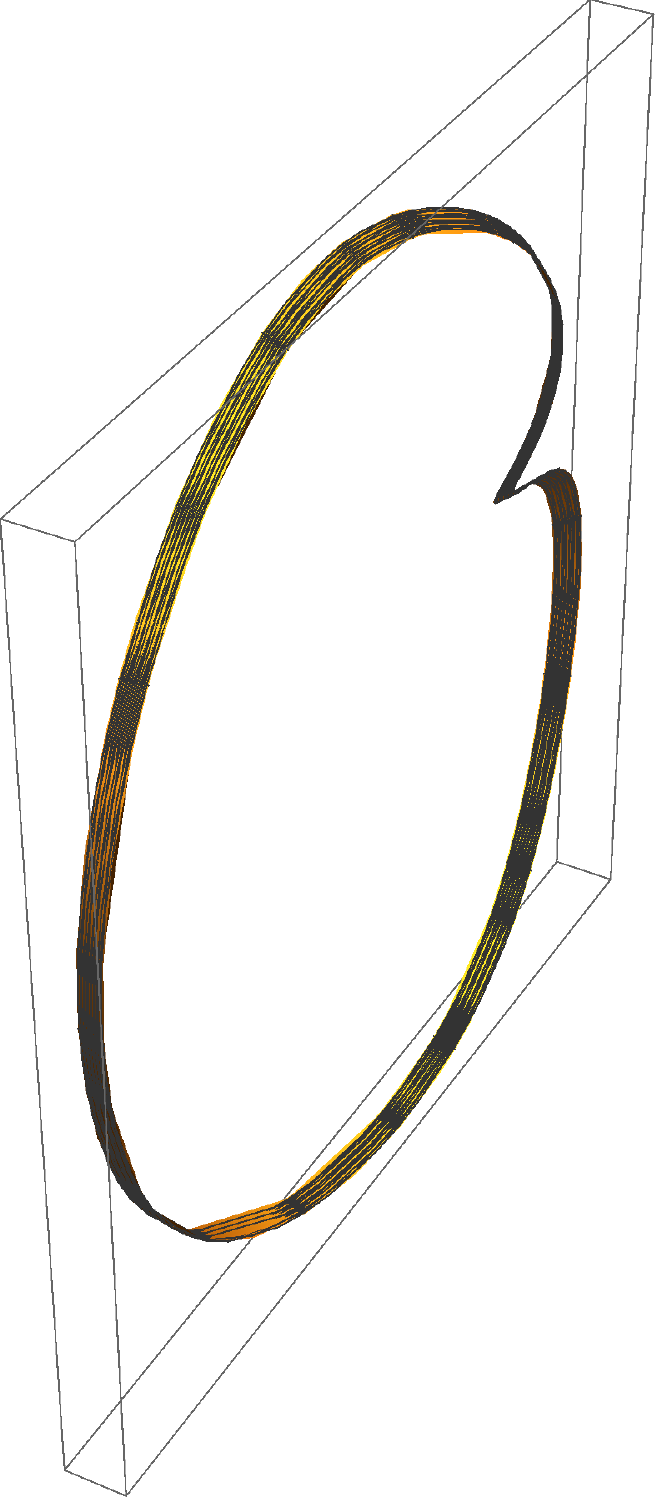}
\caption{$n=3$, $\gamma=2$, $\epsilon= 0.08$, $\tau'\equiv (4.0,4.03)$. Perturbed.}
\label{d32}
\end{minipage}
\end{figure}

\begin{figure}[h]
\begin{minipage}{12pc}
\includegraphics[width=10pc]{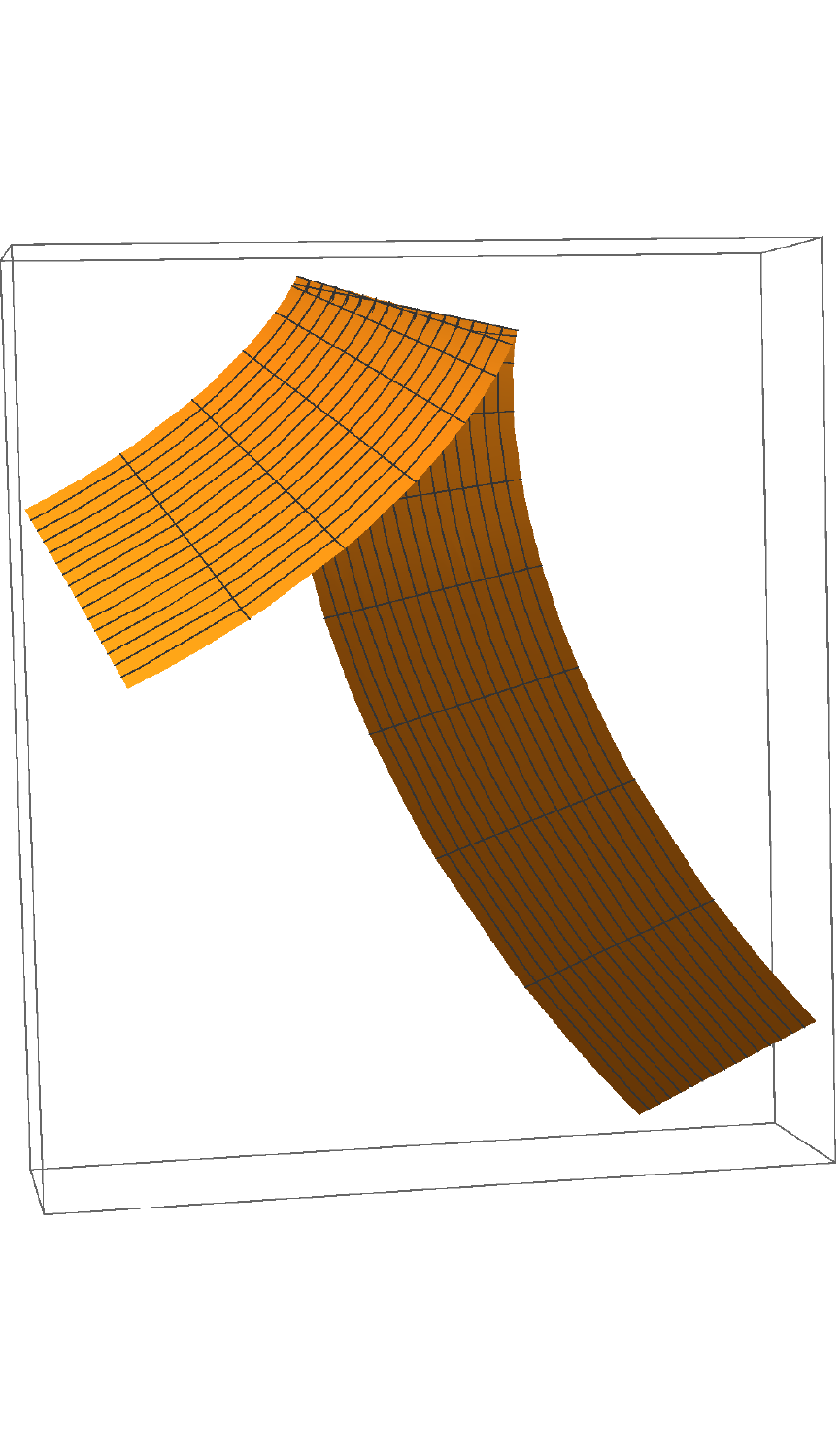}
\caption{$n=3$, $\tau'\equiv (4.0,4.03)$, unperturbed.}
\label{d41}
\end{minipage}
\hspace{1.2in}
\begin{minipage}{14pc}
\includegraphics[width=10pc]{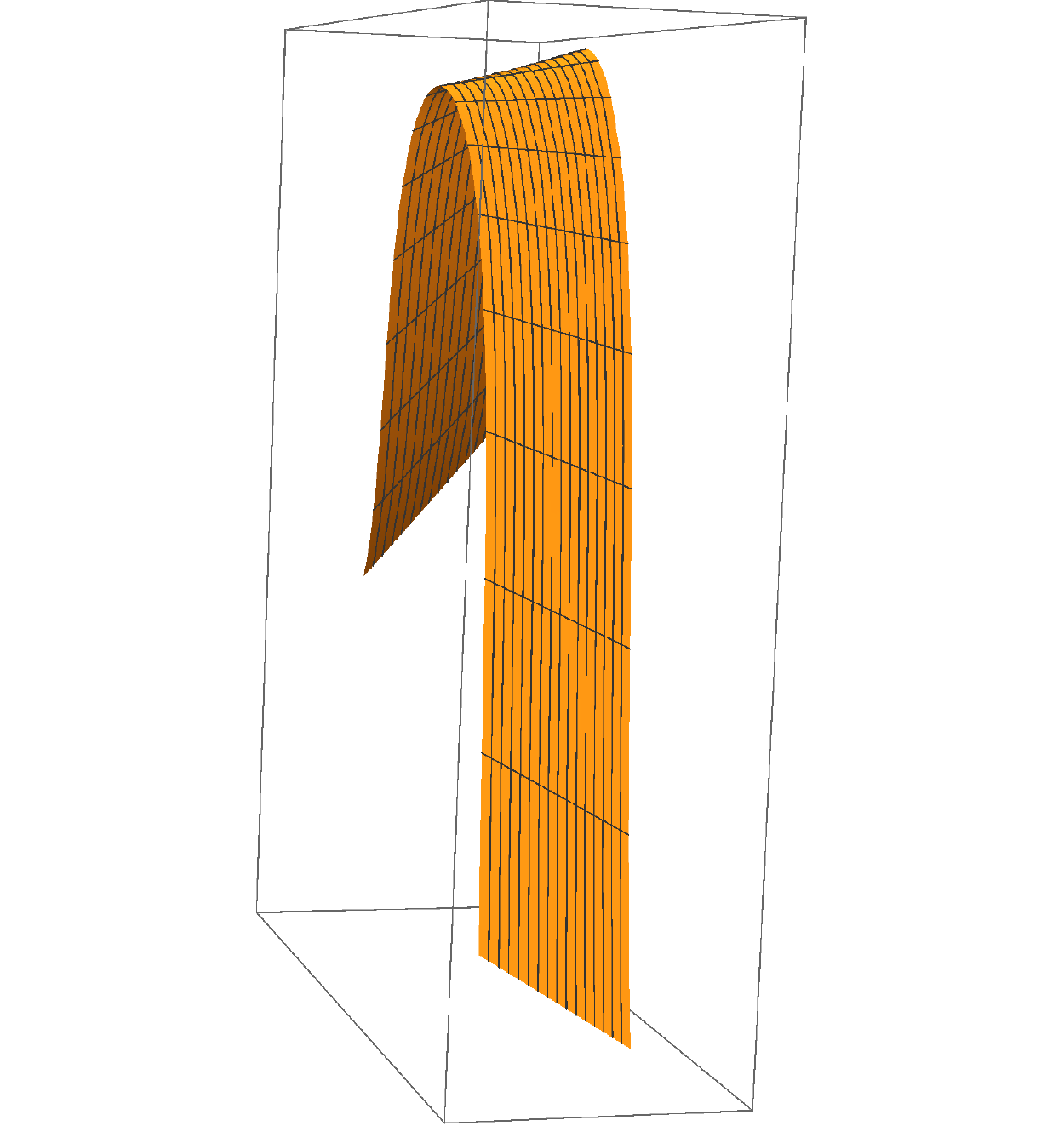}
\caption{$n=3$, $\gamma=2$, $\epsilon=0.08$, $\tau'\equiv (4.0,4.01)$, perturbed.}
\label{d42}
\end{minipage}
\end{figure}
We now proceed to find the perturbation equations. We need to
write down the tangent, normal, induced metric and extrinsic
curvature for the world sheet. The tangent vector to the
worldsheet is given as
\begin{equation}
e^{i}_\tau = \left (\dot f, \dot \rho,\dot g\right
),\hspace{0.2in} e^{i}_\sigma = \left ( 1, 0 ,\omega\right ) \ .
\end{equation}
Using the expressions for $f$, $g$ and $\rho$ quoted above the
induced line element turns out to be:
\begin{equation}
ds^2= \left (\rho^2\omega^2-1\right ) \left (-d\tau^2 +
d\sigma^2\right ) \ .
\end{equation}
The normal to the worldsheet is given as:
\begin{equation}
n^{i} = \left (\cot \omega \tau, -\frac{\bar a}{\omega \rho},
\frac{\cot \omega \tau}{\omega \rho^2}\right ) \ .
\end{equation}
The extrinsic curvature tensor turns out to be
\begin{equation}
K_{ab} = \begin{pmatrix} \bar a \omega & \omega \cr \omega & \bar
a \omega \end{pmatrix} \ .
\end{equation}
Note that all entries in the above matrix are now positive. The
quantity $K_{ab}K^{ab}$ is given as
\begin{equation}
K_{ab} K^{ab} = \frac{2\omega^2 (\bar
a^2-1)}{(\rho^2\omega^2-1)^2}= -{}^{2}R \ .
\end{equation}
The Ricci scalar of the dual worldsheet is negative (in the spiky
string case it was positive) and it diverges at the spike
locations. Substituting the above stated quantities in the
perturbation equation and separating $\tau$ and $\sigma$ dependent
parts we get
\begin{equation}
\frac{d^2 P}{d\zeta^2} +(\gamma^2 - \frac{2}{\sin^2{\zeta}})P =0 \
,
\end{equation}
where $\beta = \gamma \omega$ and $\zeta = \omega \tau$. A simple
solution for $\gamma=2$ is $P(\zeta) = \sin^2{\zeta}$. We have the
perturbation scalar as $\phi= \epsilon \rho_0 \cos 2\omega \sigma
\sin^2\omega \tau$. Following the same steps as for spiky strings,
we get, for the dual spikes,
\begin{eqnarray}
\delta t = \frac{\epsilon \rho_0}{4}\left [ \sin 2(\tau'-\sigma')+
\sin 2(n-1)(\tau'+\sigma')\right ] \ , \\
\frac{\delta\rho}{\rho} =  \frac{\epsilon n^2 \cos\left[
(n-2)\tau' + n\sigma'\right] \sin^2\left[\frac{n}{2}\tau'
+\frac{n-2}{2}\sigma'\right]}{(n-2)^2 + 4(n-1)
\sin^2\left[\frac{n}{2}\tau'+\frac{n-2}{2}\sigma'\right ]} \ ,  \\
\delta \theta = \frac{\epsilon n(n-2) \cos\left[ (n-2)\tau' +
n\sigma'\right] \sin\left[\frac{n}{2}\tau'
+\frac{n-2}{2}\sigma'\right]}{(n-2)^2+ 4(n-1)
\sin^2\left[\frac{n}{2}\tau'+\frac{n-2}{2}\sigma'\right ]}
\end{eqnarray}
Expressions for $t'$, $x'$ and $y'$ can therefore be obtained, as done earlier
for the spiky string case.

\noindent Figures 9 and 10 show the unperturbed and perturbed
worldsheets over the entire range of $\sigma'$. Figures 11 and 12
are for the unperturbed and perturbed worldsheets plotted over a
restricted range of $\sigma'$ in the neighborhood of the spike.
The rounding off of the spike is clearly visible in Figures 10 and
12. In general, solving this equation for $P(\zeta)$ we find one
oscillatory solution and one diverging solution. For the
oscillatory solution we have
\begin{equation}
P(\zeta)=\epsilon  \sin^2{\zeta}{}_{2}F_{1}(\frac{2+\gamma}{2},
\frac{2-\gamma}{2},\frac{5}{2};\sin^2{\zeta}) \ . \label{osc}
\end{equation}
It is easy to obtain various cases for different $\gamma$ values
and plot the perturbed string profile. We have shown here images
for different values of $\gamma $ too. In Figure 13, we have shown the
perturbed profile for $\gamma =4$ and in Figure 14 we have shown a closer look of the same
profile in the neighborhood of the spike, by plotting, as before,
for a restricted range of $\sigma'$. Similarly, the $\gamma = 6$
case is shown in Figure 15 and 16. In both these cases, we notice the
rounding off of the spikes.

\begin{figure}[h]
\begin{minipage}{18pc}
\includegraphics[width=18pc]{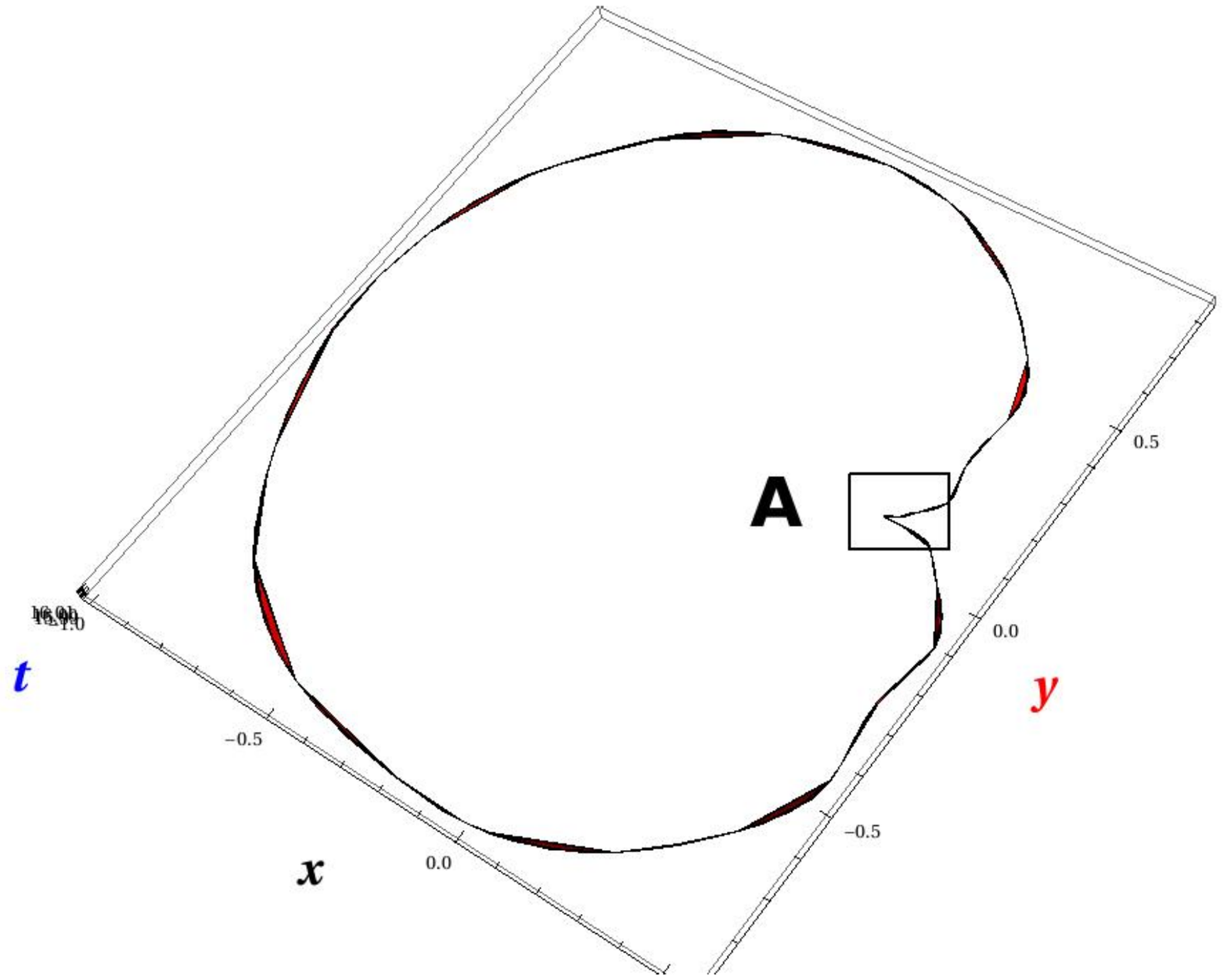}
\caption{$n=3$, $\tau'\equiv(4.0,4.0001)$, $\sigma'\equiv (0,2\pi), \gamma=4$.
}
\label{d5}
\end{minipage}
\hspace{0.7in}
\begin{minipage}{16pc}
\includegraphics[width=16pc]{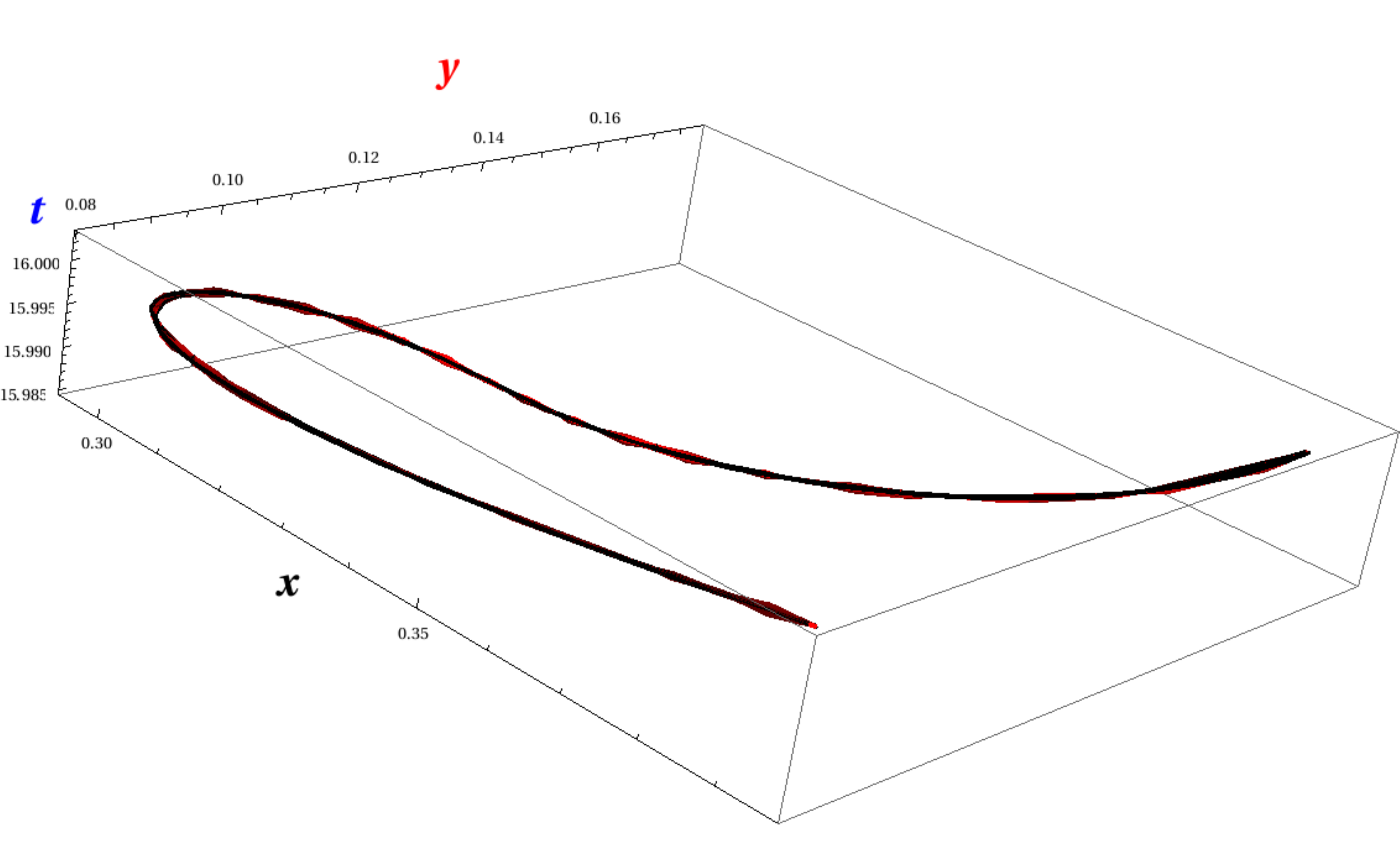}
\caption{{\bf Region A in Figure 13}, $n=3$, $\tau'\equiv(4.0,4.0001)$, $\sigma'\equiv (\pi,\frac{4\pi}{3}), \gamma=4$,
}
\label{d6}
\end{minipage}
\end{figure}
\begin{figure}[h]
\begin{minipage}{18pc}
\includegraphics[width=18pc]{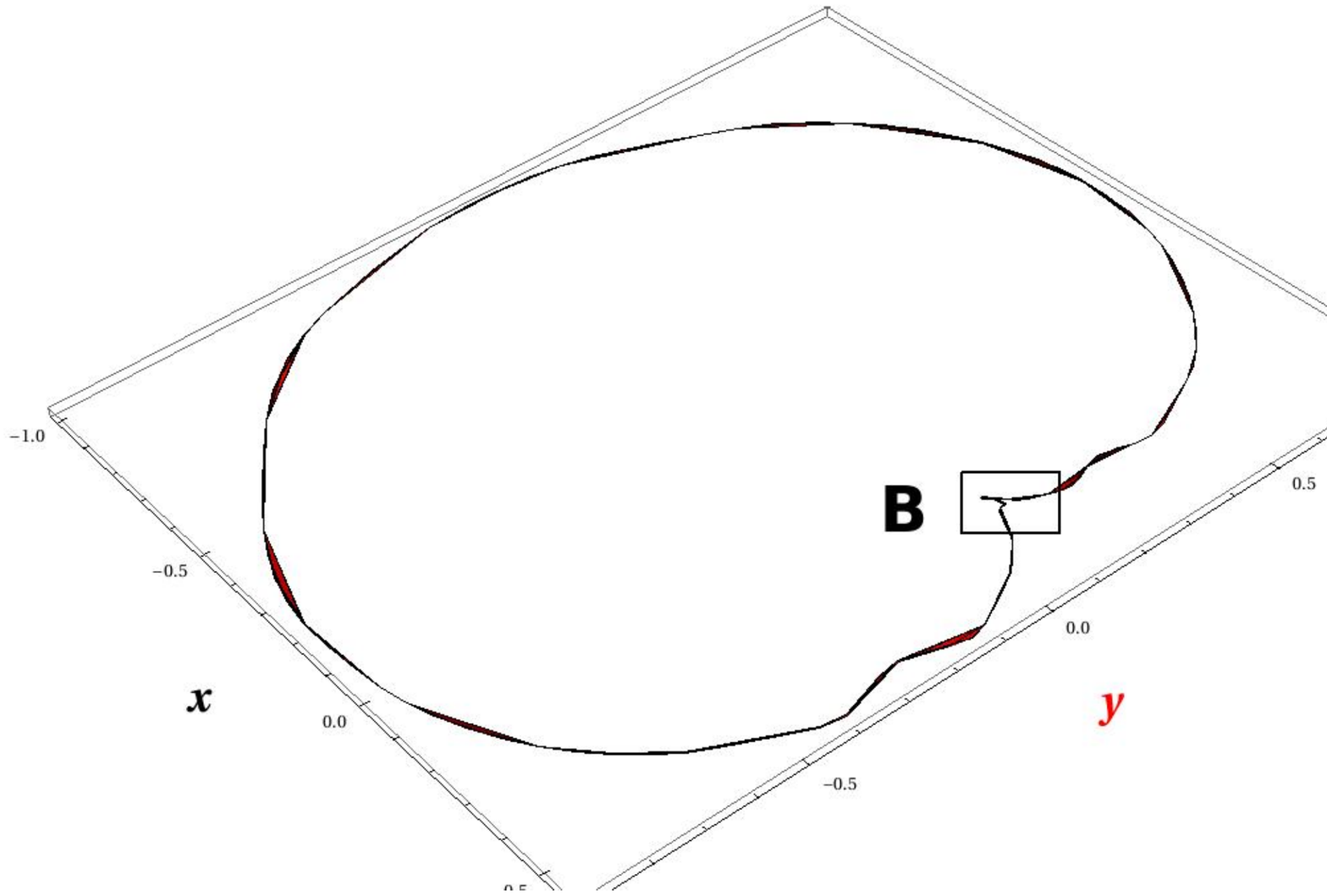}
\caption{$n=3$, $\tau'\equiv(4.0,4.0001)$, $\sigma'\equiv (0,2\pi),\gamma=6$.
}
\label{d7}
\end{minipage}
\hspace{0.7in}
\begin{minipage}{16pc}
\includegraphics[width=16pc]{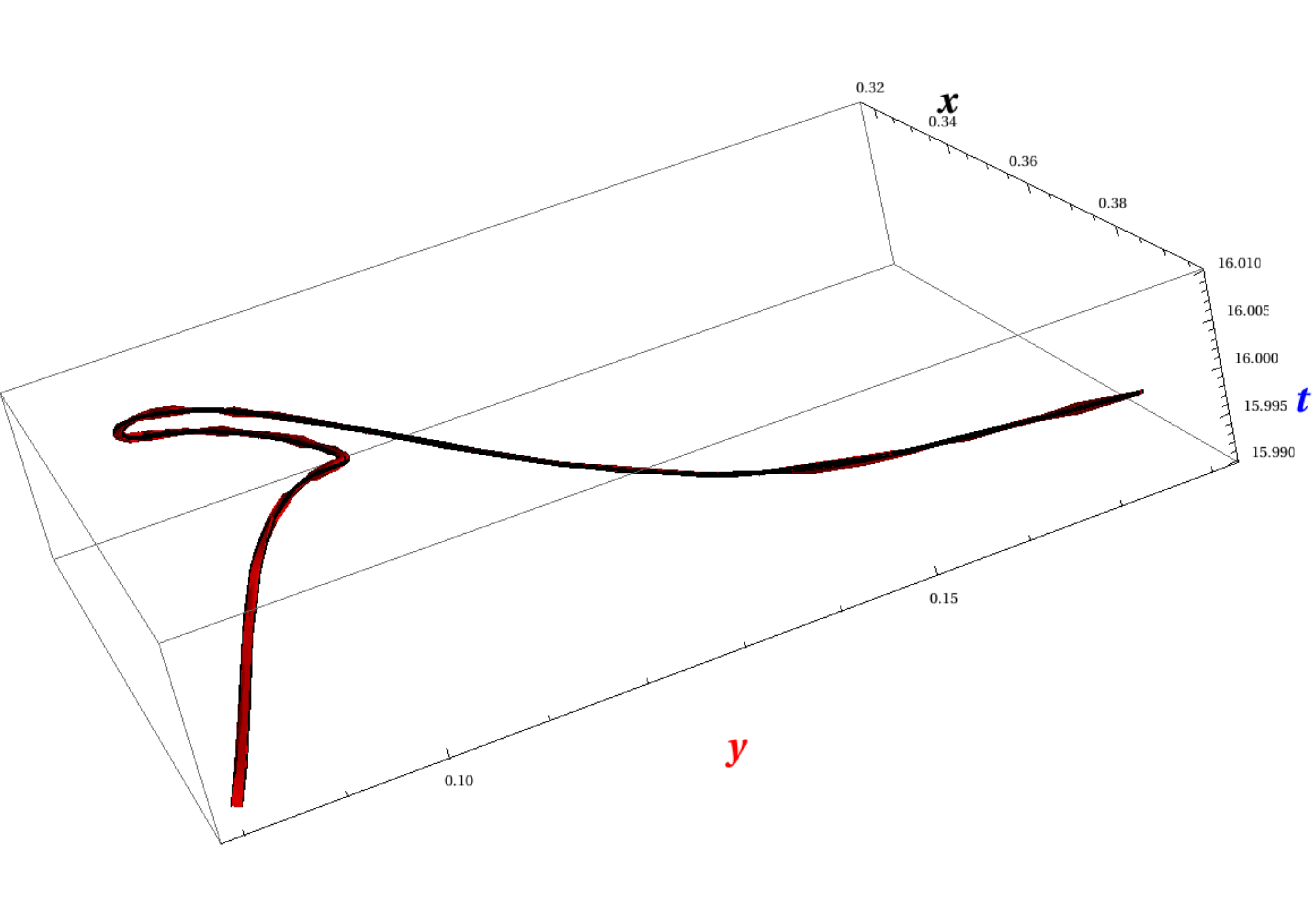}
\caption{{\bf Region B in Figure 15}, $n=3$, $\tau'\equiv(4.0,4.0001)$, $\sigma'\equiv (\pi,\frac{4\pi}{3}), \gamma=6$.
}
\label{d8}
\end{minipage}
\end{figure}
Thus, based on the above analysis, we can surely conclude that the
effect of normal perturbations of the dual spiky string also appears
through a resolution or rounding off, of the spikes. This is
indeed an indicator that both the spiky string and its dual are
stable classical solutions of the string equations of motion and
the Virasoro constraints.

\section{Extension to $3+1$ dimensions: solutions and perturbations}
It is quite straightforward to generalise the above discussion to
a $3+1$ dimensional background spacetime. We consider a flat $3+1$ dimensional
background spacetime with a line element
 \begin{equation}
ds^2= -dt^2+d\rho^2+\rho^2d\theta^2+\rho^2 \sin^2{\theta}d\phi^2 \
.
\end{equation}
We choose the embedding as
\begin{equation}
t = \tau + f(\sigma), \hspace{0.2in} \rho = \rho(\sigma),
\hspace{0.2in} \phi  =  \omega \tau + g(\sigma), \hspace{0.2in}
\theta = \frac{\pi}{2} \ .
\end{equation}
Using the above, in the string equations of motion and
constraints, we can show that the functions $\rho(\sigma)$,
$f(\sigma)$ and $g(\sigma)$ take the forms mentioned earlier. The
only novelty is the fact that we are confined to the equatorial
plane, $\theta=\frac{\pi}{2}$.

\noindent Let us now convert this solution to Cartesian coordinates. This
will enable us to plot the profiles of the unperturbed and
perturbed string worldsheets. The Cartesian embedding functions
$x$, $y$ and $z$ as functions of $\tau'$ and $\sigma'$ are given
as
\begin{eqnarray}
x &=& \frac{\rho_0}{n-2} \left [ (n-1) \cos \left
(\tau'-\sigma'\right ) - \cos \{(n-1)\left (\tau'+\sigma'\right
)\} \right ] \ ,
\nonumber \\
y &=& \frac{\rho_0}{n-2} \left [ (n-1) \sin \left (\tau'-\sigma'\right ) -
\sin \{(n-1)\left (\tau'+\sigma' \right )\} \right ]  \ . \nonumber \\
z &=& 0 \ . \nonumber \\
\end{eqnarray}
To work out the perturbations we will proceed as the $2+1$ case.
The tangent vectors to the worldsheet are,
\begin{equation}
e^{i}_\tau = \left (1,0,0,\omega\right ), \hspace{0.2in}
e^{i}_\sigma = \left ( f', \rho',0,g'\right ) \ .
\end{equation}
The induced metric is, as before.
\begin{equation}
ds^2= \left (1-\rho^2\omega^2\right ) \left (-d\tau^2 +
d\sigma^2\right ) \ .
\end{equation}
The number of the normals on any embedded surface is given by
$N-D$, where $N$ is the background spacetime dimension and $D$ is
the dimension of the surface. In the $2+1$ case, there is only one
normal but for the $3+1$ case there will be two normals. The
normals to the worldsheet are chosen as
 \begin{equation}
n^{i(1)} = \left (\tan \omega \sigma, -\frac{\bar a}{\omega
\rho}, 0, \frac{\tan \omega \sigma}{\omega\rho^2}\right ) \ .
\end{equation}
\vspace{0.2in}
\begin{equation}
n^{i(2)}= \left (0, 0, \frac{1}{\rho}, 0 \right ) \ .
\end{equation}
The two different extrinsic curvature tensors (along the two
different normals) turn out to be
\begin{equation}
K_{ab}^{(1)} = \begin{pmatrix} -\bar a \omega & -\omega \cr
-\omega & -\bar a \omega \ , \end{pmatrix}
\end{equation}
and
\begin{equation}
K_{ab}^{(2)} = 0 \ .
\end{equation}
Hence the quantity $K_{ab}^{\alpha}K_{\beta}^{ab}$ is given as
\begin{equation}
K_{ab}^{\alpha} K_{\beta}^{ab} = -\frac{2\omega^2 (1-\bar
a^2)}{(1-\rho^2\omega^2)^2} \delta^\alpha_\beta \ .
\end{equation}
Since the background is flat, the Riemann tensor is zero and
therefore the equation for the perturbation scalar along
$n^{i(1)}$ becomes:
\begin{equation}
\left ( -\frac{\partial^2}{\partial \tau^2} +
\frac{\partial^2}{\partial \sigma^2} \right )\phi^{(1)} -
\frac{2\omega^2 (1-\bar a^2)}{(1-\rho^2\omega^2)^2} \phi^{(1)} =0
\ .
\end{equation}
Using the functional form of $\rho (\sigma)$ one can reduce this
equation to
\begin{equation}
\left (- \frac{\partial^2}{\partial \tau^2} +
\frac{\partial^2}{\partial \sigma^2} \right ) \phi^{(1)} -
2\omega^2 \sec^2\omega \sigma \phi^{(1)} =0 \ . \label{peq4}
\end{equation}
Let us take a simple separable ansatz
\begin{equation}
\phi^{(1)}{(\sigma)} = e^{i\beta \tau} \chi(\sigma) \ .
\end{equation}
Using a simple transformation $\xi = \omega\sigma$ the equation
for $\chi(\sigma)$ reduces to that for the $P(\sigma)$ in the
$2+1$ case. The equation for the perturbation scalar along the
second normal turns out to be
\begin{equation}
\left (- \frac{\partial^2}{\partial \tau^2} +
\frac{\partial^2}{\partial \sigma^2} \right ) \phi^{(2)} = 0 \ .
\label{wve}
\end{equation}
The general solution of (\ref{wve}) is of the form
\begin{equation}
\phi^{(2)}(\sigma) = C_1 g(\tau+\sigma) + C_2 h(\tau- \sigma) \ .
\label{ws}
\end{equation}
The general solution for (\ref{peq4}) is
\begin{equation}
\chi(\xi) = C_1\chi_1 + C_2 \chi_2 \ ,
\end{equation}
where $\chi_1$ and $\chi_2$ are $P_1$ and $P_2$ respectively, as described
in the previous section in (\ref{s1}) and (\ref{s2}) and
$\chi_1$ is the converging solution. If we choose $\gamma= 2$
a solution is given as
\begin{equation}
\chi_1(\xi)= \epsilon \rho_0 \cos^2{\xi} \ .
\end{equation}
For the $\tau$ solution we take it's real part(i.e. $\cos{\beta
\tau} = \cos{2 \omega \tau}$). For (\ref{ws}) we can consider a
special case without loss of generality, as the following
\begin{equation}
\phi^{(2)} = {\bar \epsilon} \rho_0 \{\cos{\frac{(\tau + \sigma) \rho_0}{\rho_1}} + \cos{\frac{(\tau-\sigma)\rho_0}{\rho_1}}\}
\end{equation}
where $\bar \epsilon\neq \epsilon$ is the amplitude of the perturbation.
The perturbed embedding is given as
\begin{eqnarray}
t'= t+ \phi^{(1)} n^0_{(1)} + \phi^{(2)} n^0_{(2)} = t + \epsilon \rho_0 \cos \gamma \omega \tau \sin \omega \sigma
\cos \omega \sigma \ ,  \\
\rho' = \rho + n^1_{(1)} \phi^{(1)} +  n^1_{(2)} \phi^{(2)} = \rho -
\epsilon\rho_0\frac{\bar a \cos \gamma \omega \tau \cos^2 \omega \sigma }
{\sqrt{{\bar a}^2
+(1-{\bar a}^2) \sin^2 \omega \sigma}} \,  \\
\theta' = \theta +  n^2_{(1)} \phi^{(1)} + n^2_{(2)} \phi^{(2)}
=\frac{\pi}{2} + \frac{{\bar\epsilon} \rho_0}{\rho}
\{\cos{\frac{(\tau + \sigma) \rho_0}{\rho_1}} +
\cos{\frac{(\tau-\sigma)\rho_0}{\rho_1}}\} \ ,
\hspace{0.2in} \\
\phi'= \phi + n^3_{(1)} \phi^{(1)} + n^3_{(2)} \phi^{(2)} = \phi +
\epsilon \rho_0 \frac{\omega \cos \gamma \omega \tau \sin \omega
\sigma \cos \omega \sigma }{{\bar a}^2 +(1-{\bar a}^2) \sin^2
\omega \sigma} \ .
\end{eqnarray}
We need to switch back to Cartesian coordinates to get a clearer
picture here. Defining $x'= \rho' \sin{\theta'} \cos{\phi'}$, $y'=
\rho' \sin{\theta'} \sin{\phi'}$ and $z'= \rho' \cos{\theta'}$
and using $\rho' = \rho + \delta \rho$, $\theta' = \theta + \delta
\theta$ and $\phi' = \phi + \delta \phi$ this eventually yields
\begin{eqnarray}
x' = x + \frac{\delta \rho}{\rho} x - \delta \phi y, \>\>\>  y' =
y +\frac{\delta\rho}{\rho} y+ \delta \phi x , \>\>\>  z' = -\rho
\delta \theta \ . \nonumber \\
\end{eqnarray}
\noindent Using the worldsheet coordinate transformation
(\ref{eq1}) and the relation (\ref{eq3}) we can write the
perturbed solution in Cartesian coordinates, for $\tau'=0$  as
follows,
\begin{equation}
x' = x (1-\frac{\epsilon (n-2)^2 \cos^2{\frac{n\sigma'}{2}}
 \cos{ (n-2)\sigma'}}{(n-2)^2+4 (n-1) \sin^2{\frac{n \sigma'}{2}}})
+y \frac{\epsilon n (n-2) \sin{\frac{n \sigma'}{2}} \cos{\frac{n
\sigma'}{2}} \cos{(n-2) \sigma'} }{(n-2)^2 + 4 (n-1)
\sin^2{\frac{n \sigma'}{2}}} \ ,
\end{equation}
\begin{figure}[h]
\begin{minipage}{18pc}
\includegraphics[width=18pc]{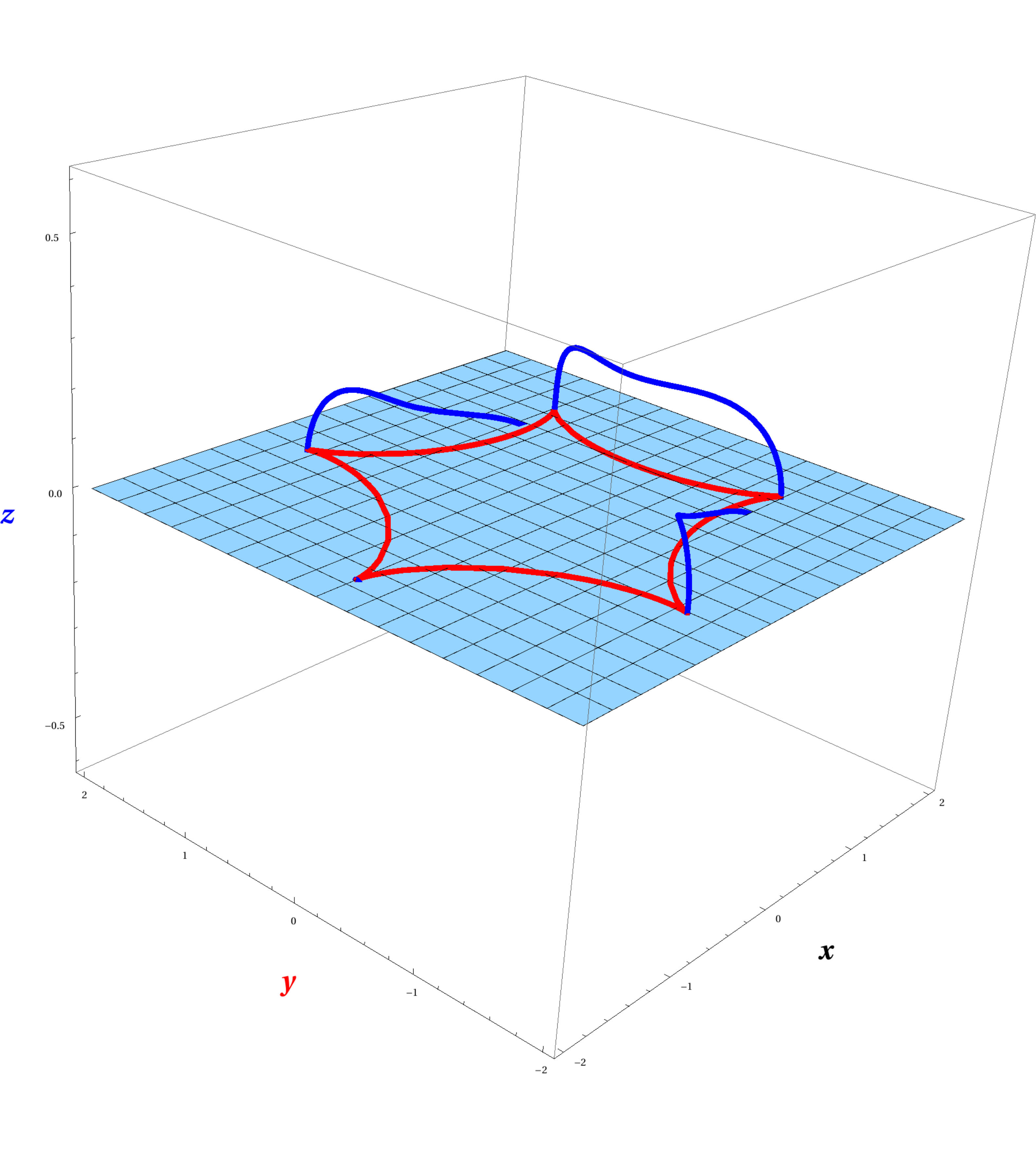}
\caption{Top view w.r.t. $z=0$ plane.}
\label{x1}
\end{minipage}
\begin{minipage}{18pc}
\includegraphics[width=18pc]{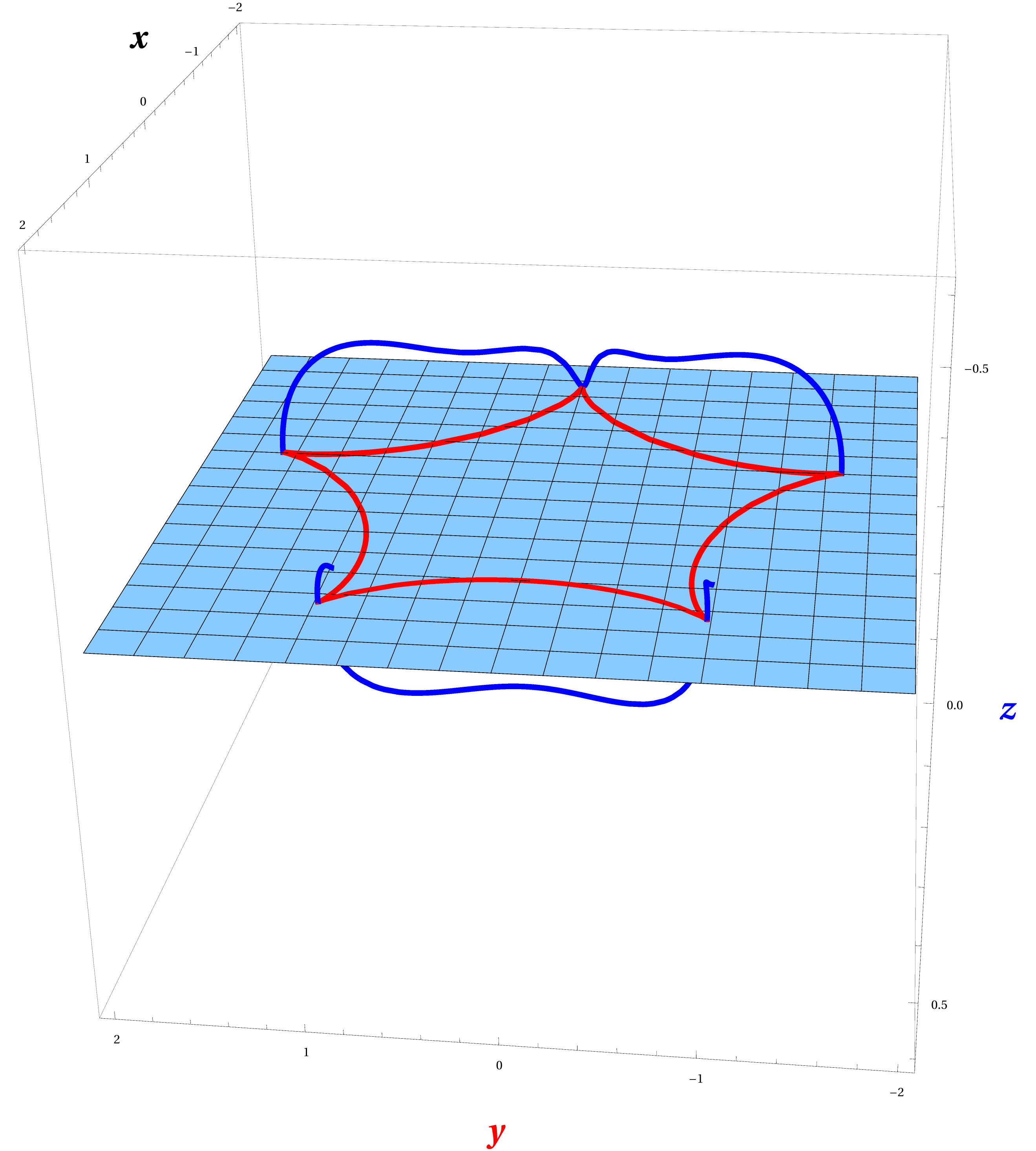}
\caption{View from below $z=0$ plane.}
\label{x2}
\end{minipage}
\begin{center}
\begin{minipage}{18pc}
\includegraphics[width=18pc]{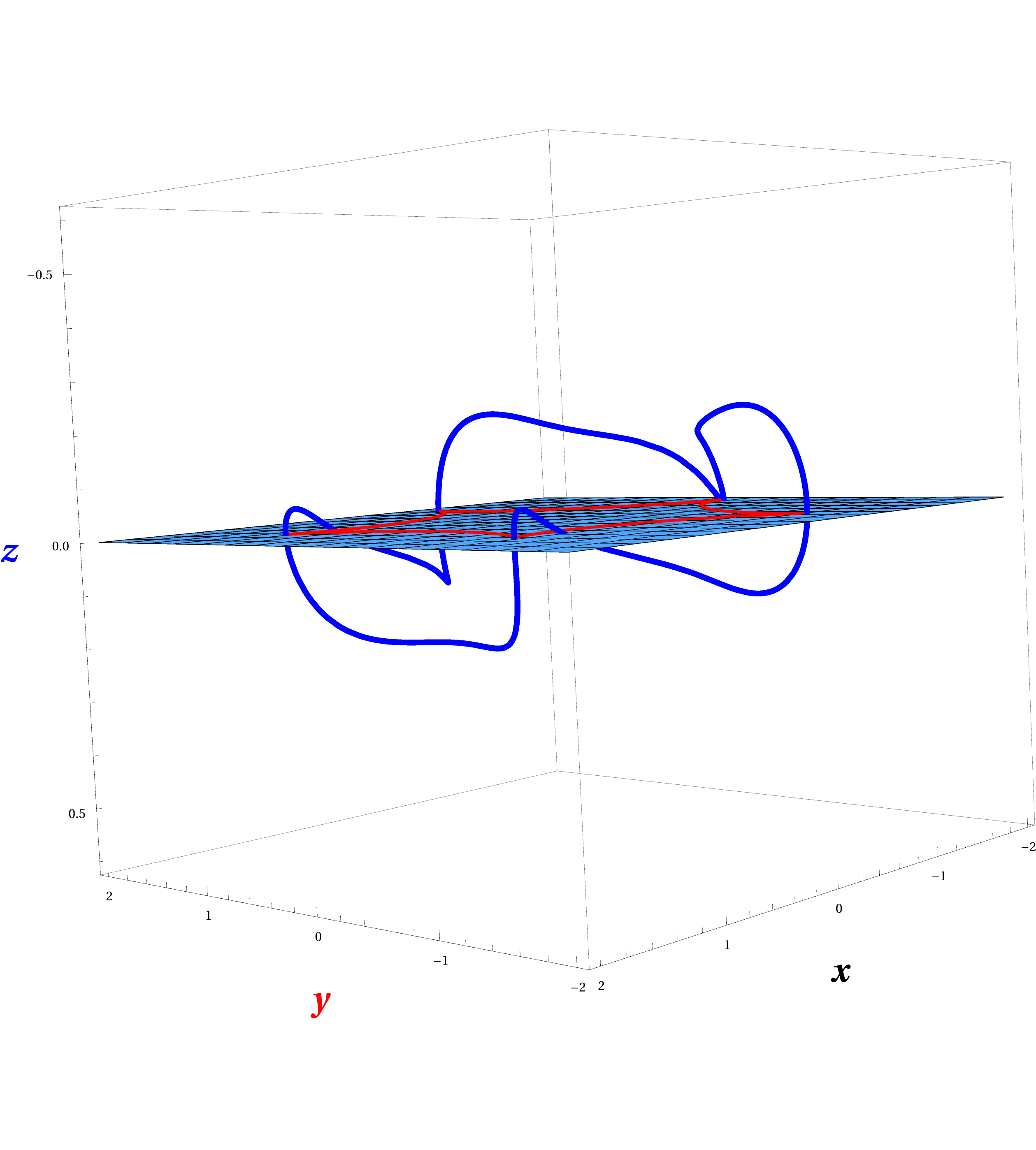}
\caption{Side view. }
\label{x3}
\end{minipage}
\end{center}
\end{figure}

\begin{equation}
y' = -x \frac{\epsilon n (n-2) \sin{\frac{n \sigma'}{2}}
\cos{\frac{n \sigma'}{2}} \cos{(n-2) \sigma'}}{(n-2)^2 + 4 (n-1)
\sin^2{\frac{n \sigma'}{2}}} + y (1-\frac{\epsilon (n-2)^2
\cos^2{\frac{n\sigma'}{2}}
 \cos{(n-2)\sigma'}}{(n-2)^2+4 (n-1) \sin^2{\frac{n
 \sigma'}{2}}}) \ , \hspace{0.2in}
\end{equation}
\begin{equation}
z' = -{ \epsilon' } \{\cos{(n-1) \sigma'} + \cos{\sigma'}\} \ .
\end{equation}

\noindent Here $\epsilon' = {\bar \epsilon} \rho_0$. In Figures
\ref{x1}, \ref{x2} and \ref{x3} we have shown the perturbed (blue) and unperturbed
(red) string profiles together assuming $\epsilon=\bar \epsilon$.
The red curve is lying on the $z=0$ plane whereas the black curve
is the  perturbed profile. In the perturbed profile we can see
that the spikes have been rounded off and the string is oriented
in three dimensions (not exclusively on the $z=0$ plane). Figures
\ref{x1} and \ref{x2} show how the perturbed string would look like from the
upper ($z>0$) and lower ($z<0$) regions of the $z$-axis. Figure \ref{x3}
combines these two pictures and demonstrates how the string is now
no longer confined in the $z=0$ plane. The feature shown in
Figures \ref{x1},\ref{x2} and \ref{x3} demonstrate how the spikes disappear and
smooth out because of the presence of the extra dimension ($z$
coordinate). We have verified this for other values of $\tau$ and
also for the dual spiky string. It is to be noted that for the
simple case of $\epsilon=0$ and $\bar\epsilon\neq 0$ we have $z'$
different from $z=0$ whereas the $x'=x$ and $y'=y$. Here too the
perturbed profile spreads into the $z$ direction leading to the
spikes disappearing. One may also consider the full worldsheet by
taking into account the $t$ (or $t'$) direction as well. Such a
scenario, of course, cannot be visualised or plotted. But the
result about rounding off of the spikes remains unaltered.

\section{Conclusions}
In this paper, we have first shown that the simplest spiky strings
in a $2+1$ dimensional flat background are stable against normal
perturbations. The stability is demonstrated through the rounding
off of the spikes when the string is slightly perturbed. We have
shown this explicitly by solving the perturbation equations and
obtaining the perturbed string configurations. The rounding off of
the spikes (for nonzero angular momentum or winding), at the level
of classical solutions, have been noticed earlier while studying
spiky strings and their duals in $AdS_3 \times S^1$ with energy
(E), spin (S) in $AdS_3$ and angular momentum (J), winding (m) in
$S^1$ \cite{Ishizeki:2008tx}\cite{mosaffa:2007}.

\noindent In our analysis here which, of course, is for spiky strings and
their duals in a flat background, it is  the worldsheet
perturbations which do the job of rounding off of the spikes. The
perturbations are stable for particular values of the parameter
$\gamma$ and also for a choice, among linearly independent
solutions (the $P_1$ solution mentioned above). Further, we have
seen that the perturbation equation turns out to be a special case
of the time independent Schrodinger equation of a particle in a
Poschl Teller potential. Our illustrations here are restricted to
some simple exact solutions of the perturbation equations
(specific values of the parameter $\gamma$) though, using the
general solution one can surely find out the nature of the
perturbation for any allowed mode (i.e. any $\gamma$).

\noindent We have also studied the dual spike solutions. In the dual spiky
string case we have first introduced a dual emdedding in the
conformal gauge (Jevicki-Jin gauge) to write down the string
configurations. Thereafter, following the same procedure as for
the spiky string case, we have obtained the perturbation equation
and its solutions. Here too, the rounding off of the spikes is
visible in the worldsheet profiles. It is worth noting here that
the perturbation equation for perturbation scalar, for the dual
spikes, is also a special case (but different from that for the
spiky string) of the Poschl Teller potential problem in quantum
mechanics. It may thus be worth addressing an inverse problem--does
there exist a string configuration for which the perturbation
equation will involve the full Poschl-Teller potential (i.e. a
combination of the two special cases we have found here).

\noindent The spiky strings in $2+1$ dimensions can easily be extended to a
$3+1$ dimensional background flat spacetime. Here we have also
worked out the normal perturbations and obtain the perturbed
worldsheet. We find that the presence of the additional dimension
in the background leads to a rounding off of the spikes. Further,
if we consider the full worldsheet evolution the rounding off
persists. Therefore, one may argue that in a $3+1$ dimensional
flat spacetime too the spiky solutions are indeed stable.

\noindent An obvious extension of this work is to investigate the
perturbations of the AdS spiky strings which are relevant in the
context of the AdS-CFT correspondence. In particular, it will be
useful to know how the perturbations are related to the operators
in the dual field theory side of the correspondence. We hope to
communicate our results on normal perturbations of the spiky
strings in AdS spacetime and also discuss possible connections
with the associated field theory, in the near future.

\section*{Acknowledgements}

\noindent The authors thank SERC, DST, Government of India for 
support through the project SR/S2/HEP-04/2011. SB would like to thank Aritra Banerjee for discussions.

\end{document}